\documentclass[12pt]{article}
\usepackage{float}
\usepackage{cite}
\usepackage{amssymb}
\usepackage{amsfonts}
\usepackage{amsmath,amsthm}
\usepackage{graphics,epsfig}
\usepackage{subfigure}
\usepackage{graphicx,epsfig, color }
 \numberwithin{equation}{section}
\begin{document}
\begin{center}
{{\bf {Thermodynamic phase transition of Anti-de Sitter
Reissner-Nordstr\"{o}m black holes with exotic Einstein-Maxwell
gravities}} \vskip 0.4 cm { Hossein Ghaffarnejad \footnote{E-mail
address: hghafarnejad@semnan.ac.ir; ORCID ID: 0000-0002-0438-6452}
and Elham Ghasemi \footnote{E-mail address:
e\_ghasemi@semnan.ac.ir}}}  \vskip 0.1 cm {\textit{Faculty of
Physics, Semnan University, P.C. 35131-19111, Semnan, Iran} }
\vskip 0.1 cm
\end{center}
\begin{abstract}
To consider the inevitable cosmic magnetic effects instead of the
unknown dark sector of matter/energy on the inflation phase of the
expanding universe some authors have proposed several extended
exotic Einstein-Maxwell gravities which are addressed in this
work. Some of these exotic models include directional interaction
terms between the electromagnetic vector field and the metric
tensor field. We use one of them to investigate the physical
effects of interaction terms on the thermodynamic behavior of the
modified Reissner-Nordstrom (RN) black hole. We use the
perturbation series method to find analytic solutions of the field
equations because of the non-linearity of the field equations
which cause they do not have analytic closed form solutions. We
investigate possibility of the Hawking-Page and the small/large
black hole phase transition and also, effects of the interaction
part of the model on possibility of the coexistence of the several
phases of the perturbed AdS RN black hole under consideration.
\end{abstract}
\section{Introduction}
Despite the consistency of Einstein's general theory of relativity
(GR) with cosmological observations \cite{1}, experiments
\cite{2}, and hence wide use of it, there are many presented
alternative gravity models which attracted so much attention
\cite{3}. The lack of comprehensiveness and validity of GR to
explain unanswered issues (e.g. dark matter/energy, the cosmic
inflation era and singularities) is the main reason for such a
replacement. In this way, modifying the GR model through unifying
the gravity with other fields such as nonlinear electrodynamics
fields is considered as an acceptable alternative theory. Born and
Infeld
 proposed nonlinear electrodynamics field as generalization of the Maxwell theory
 to re-solve the central singularity issue created by
the point  charges \cite{4}. Afterward, the Born-Infeld (BI)
theory was noticed in many respects. Hoffman studied the coupling
of Einstein`s gravity to BI theory \cite{5}. Moreover, a variety
of nonlinear electrodynamic models have been proposed and studied
in different frameworks \cite{6} such as special relativity
\cite{7}, string theory \cite{8,9,10} and notably, in cosmology to
explain the cosmic inflation era \cite{11,12} and the accelerated
expansion of the late-time universe \cite{13}. The RN black hole
as a well-known solution of the EM gravity, is a charged static
black hole with mass $M$ and charge $Q$ \cite{14,15}. In the
framework of black holes two types of black hole solutions are
obtained from lagrangian including coupling between GR and
nonlinear electrodynamics, which are generalized RN solutions
\cite{16,17,18} or the regular black hole solutions
\cite{19,20,21,22,23,24}. These solutions have been studied and
investigated in many different aspects, such as stability
properties, gravitational lensing, thermodynamic features, etc.
\cite{25,26,27,28,29,30,31,32,33,34,35}.\\ Much attention has been
drawn toward a new field of interest named quantum gravity since
Hawking discovery,
 in which  Stephen Hawking claimed that black holes also radiate \cite{36}. In this way,
 he considered  effects of quantum fields which interact with black holes
 and cause to radiate for which temperature of a radiating black hole is related to its
surface gravity \cite{37}. Moreover, Bekenstein attributed the
concept of entropy to black holes which is proportional to the
black hole's horizon surface area \cite{38}. In this meantime,
Bardeen, Carter and Hawking extracted four laws of black hole
mechanics according to the known four laws of ordinary
thermodynamics as a result of black holes behaving as
thermodynamic objects \cite{39}. Since then, people have studied
and investigated the thermodynamic properties of black holes and
their phase behaviors in many literature \cite{40,41}. For a more
detailed study of black holes thermodynamics, the extended phase
space as a modification of black hole thermodynamics has been
presented\cite{42,43,44}. This is done because of the presence of
a pressure thermodynamic variable in the black hole equation of
state. In fact, the pressure term comes from a negative
cosmological parameter originating from AdS/CFT correspondence
\cite{witten1, witten2}. In the presence of this additional term,
many of the studied black holes exhibit two different phase
transitions as;  (a) the small/large black hole phase transition
which is similar to a Van der Waals gas/fluid phase transition in
the ordinary thermodynamic systems and (b) the Hawking-Page phase
transition where an evaporating black hole completely disappears
into an ideal gas. To see thermodynamic properties and phase
transitions of such black holes  one can follow
\cite{45,46,47,48,49,50} in GR gravity, and
\cite{51,52,53,54,55,56,57,58,59,60,61,62,63,64,35,67,68,69,70,Chamblin1,Chamblin2,66}
for modified gravity theories. In this work we investigate the
physical effects of the non-minimal interaction part between the
gravity and the electromagnetic field, in studying the
thermodynamic behavior of a modified RN AdS black hole. This work
is managed as follows.\\ In the second section of this paper, we
introduce a modified Einstein-Maxwell gravity that includes a
non-minimal directional coupling term between gravity and
electromagnetic tensor field in the presence of the cosmological
constant.  Also, we obtain the corresponding Einstein metric field
equation and Maxwell electromagnetic field equations. In the third
section, we find the metric field of a spherically symmetric
static dS/AdS black hole which turned out to be a modified RN
solution. We calculate the equation of state and the Hawking
temperature of the obtained metric field in the fourth section .
In the fifth section, we investigate the possibility of the phase
transitions for different values of the parameters of the system.
The last part of the paper is dedicated to the concluding remark.
 \section{The gravity model}
Since the influence of the cosmic magnetic field on gravitational
systems is inevitable, the author of \cite{MS} proposed several
alternative models instead of the well-known EM gravity model in
order to study the impact of this field. These models were
proposed to explore how the cosmic magnetic field affects the
inflation of the universe.
  His models are
suitable because the magnetic field intensity does not suppress
during the inflation and so can support violations in the cosmic
energy density components. In the ordinary well known EM gravity
model the cosmic magnetic field suppresses suddenly and so becomes
inefficient. We choose one of his proposed alternatives to obtain
a black hole metric solution, such that
\begin{equation}\label{21}
I=\int d^4x\sqrt{g}\left[R-\Lambda-\frac{1}{4}F_{\mu\nu}F^{\mu\nu}
+\alpha F_{\rho\mu}R^{\mu}_{\eta}F^{\eta\rho} \right]
\end{equation}
where $F_{\mu\nu}=\partial_\mu A_\nu-\partial_\nu A_\mu$ is the
Maxwell antisymmetric tensor field. $A_{\mu}$ is four-vector
Maxwell potential field.  The parameter of $\alpha$ is the
interaction coupling constant between electromagnetic tensor field
$F_{\mu\nu}$ and the Ricci tensor field $R_{\mu\nu}$. Its
dimension is $(length)^2$ in the units $\hbar=G=c=1.$ $\Lambda$
represents the cosmological constant and its support is needed as
the pressure of AdS space affecting on the resulting black hole.
It is obvious that the above action functional remains un-changed
under the gauge transformation $A_{\mu} \to A_{\mu}+\partial_{\mu}
\xi$ where $\xi$ is a free scalar gauge field. In fact, this is
because $F_{\mu\nu}\to F_{\mu\nu}$ under this gauge
transformation. As an application of this model, one of the
authors of this work studied stabilization of a spherical perfect
fluid star in presence of magnetic monopole field recently
\cite{Ghaf2023}. Variation of the above action functional for the
metric tensor $g_{\mu\nu}$ yields the Einstein field equations
such that
\begin{equation}\label{Ein}
G_{\mu\nu}=\Lambda g_{\mu\nu}+T^{EM}_{\mu\nu}-\alpha
\Theta_{\mu\nu},
\end{equation}
where
\begin{equation}\label{23}
G_{\mu\nu}=R_{\mu\nu}-\frac{1}{2}g_{\mu\nu}R,
\end{equation}
is the Einstein tensor, $R_{\mu\nu}$ is the Ricci tensor,
$R=R^\epsilon_{~\epsilon}$ is the Ricci scalar  and
\begin{equation}\label{24}
T^{EM}_{\mu\nu}=-\frac{1}{4}\left[F_{\mu\alpha}F^{~\alpha}_\nu+F_{\beta\nu}F^\beta_{~\mu}
-\frac{1}{2}g_{\mu\nu}F_{\alpha\beta}F^{\alpha\beta}\right]
\end{equation}
is traceless stress tensor of the electromagnetic tensor field
$F_{\mu\nu}.$ Furthermore,

\begin{align}\label{Ttheta}\Theta_{\mu\nu}=&-\frac{1}{2}g_{\mu\nu} F_{\rho\alpha}
R^\alpha_\eta F^{\eta\rho}+ F_{\rho\mu}
R_{\nu\eta}F^{\eta\rho}+F_{\rho\eta} R^\eta_\mu F^\rho_\nu
+F_{\mu\rho} R^\rho_\eta F^\eta_\nu
\notag\\&-\frac{1}{2\sqrt{g}}\partial_\alpha\left[\nabla_\eta\left(\sqrt{g}F^\alpha_\rho
F^{\eta\rho}\right)\right]g_{\mu\nu}
+\frac{1}{\sqrt{g}}\nabla_\mu\left(\sqrt{g}F^\alpha_\rho
F^{\eta\rho}\right)\Gamma_{\nu\alpha\eta}
\notag\\&+\frac{1}{4\sqrt{g}}g_{\eta\mu}g_{\sigma\nu}\partial_\lambda\left[\nabla_\alpha\left(\sqrt{g}F^\lambda_\rho
F^{\eta\rho}\right)g^{\alpha\sigma}\right]
\notag\\&+\frac{1}{4\sqrt{g}}g_{\sigma\mu}g_{\lambda\nu}\partial_\eta\left[\nabla_\alpha\left(\sqrt{g}F^\lambda_\rho
F^{\eta\rho}\right)g^{\alpha\sigma}\right]\notag\\&-\frac{1}{4\sqrt{g}}g_{\lambda\mu}g_{\eta\nu}\partial_\sigma\left[\nabla_\alpha\left(\sqrt{g}
F^\lambda_\rho F^{\eta\rho}\right)g^{\alpha\sigma}\right]
\end{align}
is the interaction part of the stress tensor.  Through variation
of the action functional (\ref{21}) with respect to $A^\mu$, the
covariant form of the Maxwell equation is obtained such that
\begin{equation}\label{26}
\nabla_{\nu}F^{\mu\nu}=2\alpha J^{\mu},
\end{equation}
where the four-vector current density $J^{\mu}$ is defined by
\begin{equation}\label{27}
J^{\mu}=\nabla_\rho\left(R^\rho_\eta F^{\eta\mu}
\right)-\nabla_\lambda\left(R^\mu_\eta F^{\eta\lambda} \right).
\end{equation}
It is easy to show that the modified Maxwell equation above can be
shown with a covariantly  conserved form, such that
\begin{equation}\label{ModM}
    \nabla_{\nu}\Sigma^{\mu\nu}=\frac{1}{\sqrt{-g}}\partial_{\nu}\left( \sqrt{-g}\Sigma^{\mu\nu}\right) =0
    \end{equation}
by defining
\begin{equation}\label{MM}
    \Sigma^{\mu\nu}=F^{\mu\nu}-2\alpha R^{\nu}_{\eta}F^{\eta\mu}+2\alpha
    R^{\mu}_{\eta}F^{\eta\nu}.
    \end{equation}
In the next section, we investigate a spherically symmetric static
metric solution for the proposed gravity model above.
\section{Modified RN-AdS black hole metric}
Let's start with the line element of a
 spherically symmetric static curved spacetime with the following general form
\begin{equation}\label{31}
  ds^2=-A(r)dt^2+B(r)dr^2+r^2(d\theta ^2+\sin ^2\theta d\phi
  ^2).
\end{equation}
For this line element we have just t-r components for the four
vector potential as
\begin{equation}A_{\mu}(r)=(A_t(r),A_r(r),0,0)\end{equation} for which the Maxwell equation
(\ref{ModM}) reads
\begin{align}\label{max}\Sigma^{tr}(r)=\frac{Q
}{r^2\sqrt{AB}}\end{align} which by regarding the definition
(\ref{MM}) can be rewritten as
\begin{equation}\label{max1}E(r)=F_{rt}(r)=\frac{Q\sqrt{AB}
}{r^2[1+2\alpha\Omega(r)]}\end{equation} where the constant of
integration $2Q$ is related to the monopole electric charge,
$E(r)$ is its radial electric vector field and $\Omega$ is defined
by the Ricci tensor components, such that
\begin{align}\label{omega}\Omega=R^t_t+R^r_r=-\frac{A''}{AB}+\frac{{A^{\prime}}^2 }{2A^2B}+
    \frac{A^{\prime}B^{\prime}}{2AB^2} -\frac{A^{\prime}}{rAB}+\frac{B^{\prime}}{rB^2}.\end{align} The Einstein
tensor components are
\begin{equation}\label{A1}
    G^t_t=\frac{B^{\prime} }{r B^{2}}+\frac{1}{r^{2}}-\frac{1}{r^{2} B},
\end{equation}
\begin{equation}\label{A2}
    G^r_r=-\frac{A^{\prime}}{r AB}+\frac{1}{r^{2}}-\frac{1}{r^{2}B},
\end{equation}
\begin{align}\label{A3}
        G^{\theta}_{\theta}=G^{\phi}_{\phi}=&-\frac{A^{\prime\prime}}{2 AB}+\frac{{A^{\prime}}^2 }{4 A^{2}B}+\frac{A^{\prime}B^{\prime}}{4A B^{2} }
          -\frac{ A^{\prime} }{2r AB}
        +\frac{B^{\prime} }{2r B^{2}}.
\end{align}
        and the Maxwell stress tensor (\ref{24}) reads
\begin{align}\label{A4}
    &T^t_{t~EM}=T^r_{r~EM}=\frac{E^2}{4AB}\notag\\&
    T^{\theta}_{\theta~ EM}=T^{\phi}_{\phi~EM}=-\frac{E^2}{4AB}.
\end{align} Furthermore
the non-vanishing components of the stress tensor (\ref{Ttheta})
reduce to the following forms.
\begin{align}\label{A5}
    \Theta^t_t=&-\frac{E^2\Omega }{2AB}+\frac{E^2 R^r_r}{AB}-\frac{E^2}{4r^2AB}
    +\frac{1}{2r^2\sqrt{AB}}\left[\left(\frac{\omega}{AB^2} \right)''+\left(\frac{\omega B^{\prime}}{AB^3}\right)'   \right]
    \notag\\&
    +\frac{A}{4r^2\sqrt{AB}}\left[ \frac{1}{B}\left(\frac{\omega}{A^2B}\right)''-\frac{B^{\prime}}{B^2}\left(\frac{\omega}{A^2B} \right)'
     +\left( \frac{\omega A^{\prime}}{A^3B^2} \right)' \right],
    \end{align}
   \begin{align}\label{A6}
      \Theta^r_r=&-\frac{E^2\Omega }{2AB} +\frac{E^2R^t_t}{A}-\frac{E^2}{4r^2AB}
      -\frac{E^2}{2AB^2}\left[\left( \frac{A^{\prime}}{A}\right)^2 +\left(\frac{B^{\prime}}{B} \right)^2  \right]
        \notag\\&
        +\frac{1}{2r^2\sqrt{AB}}\left[ \left(\frac{\omega}{AB^2} \right)'' +\left(\frac{\omega B^{\prime}}{AB^3} \right)' \right]
        \notag\\&
    -\frac{1}{2r^2B\sqrt{AB}}\left[ A^{\prime}\left(\frac{\omega}{A^2 B} \right)'+B^{\prime} \left(\frac{\omega}{AB^2} \right)' \right]
    \notag\\&
    -\frac{B}{4r^2\sqrt{AB}}\left[\frac{1}{B}\left(\frac{\omega}{AB^2} \right)''-\frac{B^{\prime}}{B^2}\left(\frac{\omega}{AB^2} \right)'
    +\left(\frac{\omega B^{\prime}}{AB^4} \right)'   \right]  ,
    \end{align}
    \begin{align}\label{A7}
        \Theta^{\theta}_{\theta}=\Theta^{\phi}_{\phi}=-\frac{E^2\Omega}{2AB}+\frac{1}{2r^2\sqrt{AB}}
        \left[\left(\frac{\omega}{AB^2} \right)''+\left(\frac{\omega B^{\prime}}{AB^3} \right)'  \right],
    \end{align}
where we defined
\begin{equation}\omega(r)=r^2E^2\sqrt{AB}.\end{equation}
 The
radial electric field is defined by $E(r)=F_{rt}=A_t^{\prime}(r)$
for which $^{\prime}$ denotes to the derivative with respect to
the radial coordinate $r$. It is easy to check that in the absence
of the cosmological constant and interaction terms, i.e.,
$\Lambda=0$ and $\alpha=0$, the Einstein metric equation
(\ref{Ein}) has a simple analytic closed form of the metric
solution so called as the RN black hole geometry. In the presence
of the additional $\alpha$ term, the dynamical field equations
have complicated forms and do not have exact analytic solution.
Hence, we use the perturbation series method to solve the field
equations, because we need an analytic exact form for the equation
of state of a modified RN AdS black hole. This is found usually by
solving the equation of the black hole event horizon. Thus we
first should obtain analytic form of metric solution of a modified
RN black hole in the presence of the $\alpha$ dependent
interaction term. We assume that this additional interaction term
behaves as perturbation for which $\alpha$ should have small
values. Then we substitute
 the perturbation series of the fields
\begin{align}\label{ser}&A(r)=A_0(r)+\alpha A_1(r)+O(\alpha^2)\notag\\&B(r)=B_0(r)+\alpha
B_1(r)+O(\alpha^2)\notag\\& E(r)=E_0(r)+\alpha E_1(r)+O(\alpha^2)
\end{align}
 into the electric field (\ref{max1}) and into the Einstein metric equation
(\ref{Ein}). To find the order solutions of the fields $\{A_i,B_i,
E_i\}$ We must solve the field equations order by order which we
do in the next subsections.
\subsection{Zero order solutions.} Substituting the series
expansions (\ref{ser}) into the electric field (\ref{max1}), also,
into $tt,$ $rr$ and $\theta\theta$ components of the Einstein
metric equation (\ref{Ein}) we find the following relation
respectively.
\begin{align}\label{315}&E_0(r)=\frac{Q}{r^2}\end{align}\begin{align}\label{316}&{\frac {B_0' }{r B_0^2}}+\frac{1}{{r}^{2}}-{\frac {1}{{r}^{2}{ B_0}
 }}- \frac{E_0^2}{4A_0B_0}-\Lambda =0\notag\\&-{\frac {A_0' }{r A_0 B_0}}+\frac{1}{{r}^{2}}-{\frac {1}{{r}^{2}{ B_0}
 }}- \frac{E_0^2}{4A_0B_0}-\Lambda =0
\end{align}\begin{align}\label{317}&-\,{\frac {A_0''}{2{A_0} {B_0}  } }+\,{\frac {  {A'_0}^2}{ 4 {A_0}^{2}{ B_0}
}}+\frac{A'_0B'_0}{4A_0B_0^2}-\frac{A'_0}{2rA_0B_0}+\frac{B'_0}{2rB_0^2}
+\frac{E_0^2}{4A_0B_0}-\Lambda =0
\end{align}
which should be solved synchronously. It is easy to show that the
two  equations  (\ref{316}) reduce to the following condition.
\begin{equation}\label{318}A_0(r)B_0(r)\equiv constant=1.\end{equation}
This implies that the above differential equations are not
independent and they can only determine one of the fields $A_0(r)$
and $B_0(r).$ Hence we need other independent equation to
determine both of these fields. A further independent equation is
actually (3.17) which by substituting (\ref{315}) and (\ref{318})
reads to the following form.
\begin{equation}\frac{A_0''}{2}+\frac{A'_0}{r}-\bigg({\frac {{Q}^{2}}{{r}^{4}}}-\Lambda\bigg)=0
\end{equation}
with the solution
\begin{equation}A_0(r)=1-\frac{2M}{r}+\frac{Q^2}{r^2}-\frac{\Lambda}{3}r^2\end{equation} in which the constant of integration $2M$ is assumed to be two times of the black
hole`s ADM mass. By having the obtained zero order solution above
we investigate the linear order solution of the fields now.
\subsection{Solutions in order of $O(\alpha)$}

Substituting the zero order solutions above into the linear part
of $tt,$ $rr$ and $\theta\theta$ components of the Einstein field
equations, the following relations are resulted respectively.
\begin{equation}E_1(r)=\frac{4Q}{r^2}\bigg(\frac{Q^2}{r^4}-\Lambda\bigg)\end{equation}
\begin{align}\label{Gt1}f_1B_1+\frac{A_0^2B_1^\prime}{r}+\frac{E_0^2A1}{4A_0}-\frac{E_0E_1}{2}+f_2=0\end{align}\begin{equation}\label{Gr1}f_3B_1+f_4A_1 -\frac{E_0E_1}{2}-\frac{A'_1}{r}+f_5=0
\end{equation}
and
\begin{align}\label{Gthet1}&-\frac{A''_1}{2}+ f_6 A'_1 + f_7B'_1 +f_8 { B_1}  +f_9 { A_1}+f_{10} =0
\end{align}
where $f_i(r); i=1,2,\cdots10,$ are defined in terms of the zero
order of the solutions and presented in the appendix 1. To find
$A_1(r)$ and $B_1(r)$ in fact, we need only two equations of the
relations (\ref{Gt1}), (\ref{Gr1}) or (\ref{Gthet1}) and so one of
them is a constraint condition between the solutions. Hence in
order to solve these equations we first eliminate $B_1(r)$ and
$B'_1(r)$ between the equations (\ref{Gt1}) and (\ref{Gthet1})
such that
\begin{align}B_1=\Sigma_1(r),~~~B'_1(r)=\Sigma_2(r)\end{align} with \begin{equation}\label{lnB1}\ln B_1(r)=\int \bigg(\frac{\Sigma_2}{\Sigma_1}\bigg)dr\end{equation}in which \begin{align}&\Sigma_1=\frac{\frac{E_0E_1f_7}{2}-f_2f_7+\frac{A_0^2f_{10}}{r}-\frac{A_0^2
A''_1}{2r}+\frac{f_6A_0^2A'_1}{r}+\big(\frac{f_9A_0^2}{r}-\frac{f_7E_0^2}{4A_0}\big)A_1}{f_1f_7-\frac{A_0^2f_8}{r}}\notag\\&
\Sigma_2=\frac{-\frac{E_0E_1f_8}{2}+f_2f_8-f_1f_{10}+\frac{f_1A''_1}{2}-f_1f_6A_1^\prime+\big(\frac{f_8E_0^2}{4A_0}-f_1f_9\big)A_1}{f_1f_7-\frac{A_0^2f_8}{r}}.\end{align}
If we have an explicit form of the function $A_1(r)$ then
$\Sigma_{1,2}$ are determined and we can calculate the integral
equation (\ref{lnB1}), but $A_1$ is determined by solving the
equation (\ref{Gr1}) when we substitute $B_1=\Sigma_1$ such that
\begin{align}\label{329}A''_1+X(r)A'_1+Y(r)A_1=Z(r)\end{align} where we defined \begin{align}&X=\frac{2f_1f_7}{f_3A_0^2}-2f_6-\frac{2f_8}{rf_3}\notag\\
&Y=\frac{rf_7E_0^2}{2A_0}+\frac{2f_8f_4}{f_3}-2f_9-\frac{2rf_1f_4f_7}{f_3A_0^2}
\notag\\&Z=\frac{rf_7}{A_0^2}\bigg[E_0E_1\bigg(1-\frac{f_1}{f_3}\bigg)+\frac{2f_1f_5}{f_3}-2f_2\bigg]+rf_{10}-\frac{2f_8f_5}{f_3}+\frac{f_8E_0E_1}{f_3}.\end{align}
Substituting the definitions (\ref{def}) into the relations above
we find \begin{align}\label{XYZ}&X=\frac{4}{r}\bigg(\,{\frac
{3\,{Q}^{2}-2\,{r}^{2}}{ 4\,\Lambda\,{r}^{4}+3\,{Q}^
{2}-4\,{r}^{2}  }}\bigg) \notag\\&Y=\frac{F_1(r)}{12r^5(4\Lambda
r^4+3Q^2-4r^2)(\Lambda
r^4+6Mr-3Q^2-3r^2)^2}\notag\\&Z=\frac{Q^2F_2(r)}{6r^8(4\Lambda
r^4+3Q^2-4r^2)(\Lambda r^4+6Mr-3Q^2-3r^2)^2}\end{align} in which
explicit forms of the functions $F_{1,2}(r)$ are given in the
appendix 2.\\
 Asymptotic behavior of the functions (\ref{XYZ}) vanish  at infinity $r\to\infty$ but reduce to the following forms in limits $r\to0.$  \begin{align}\label{coef}&\left(%
\begin{array}{c}
  X_0(r) \\
  Y_0(r) \\
  Z_0(r) \\
\end{array}%
\right)=\lim_{r\to0}\left(%
\begin{array}{c}
  X(r) \\
  Y(r) \\
  Z(r) \\
\end{array}%
\right)\sim\left(%
\begin{array}{c}
  4/r \\
  -MQ^2/4r^5 \\
  -14Q^4/r^8\\
\end{array}%
\right).\end{align} Regarding the asymptotic behavior of the
coefficients (\ref{coef}) one can find general solutions of the
equation (\ref{329}) which are given by the first order Bessel
$J_1(1/r^\frac{3}{2})$ and Newmann $N_1(1/r^\frac{3}{2})$
functions, such that
\begin{align}A_1(r)\sim{\frac {{C_2}}{\tau}{{ J}_{1}\left(\,{\frac {1}{\tau}}\right)}}+{\frac {{C_1}}{\tau}{{ N}_{1
}\left(\,{\frac{1}{\tau}}\right)}}-\frac{ 504}{M^2\tau^{2}}
\end{align}
 with \begin{equation}\label{336}\tau=\frac{3r^{\frac{3}{2}}}{Q\sqrt{M}}\end{equation} where $C_{1,2}$ are constants that need to be determined by suitable initial
 conditions. It is obvious that the dimensionless parameter $\tau$
 takes positive numeric values for $Q>0$ but it takes  negative numeric
 values for $Q<0.$ We will use  two different regions called small black holes with $|\tau|<1$ and large black
 holes $|\tau|>1$ in what follows.
  It is useful to find asymptotic behavior of the solution above near the point $\tau=1,$ such that
   \begin{align}\label{337}&A_1(\tau)\sim0.44C_2- 0.78C_1-\frac{504.0}{{M}^{2}}- \bigg(0.77C_2+0.088C_1\notag\\&~~~~~~-\frac{1008.0}{{M}^{2}}\bigg)
 \left(t- 1\right)+O((\tau-1)^2),~~~\tau<1\notag\\&
 ~~~~~~\sim-\frac{C_1}{\pi}\frac{\ln(2\tau)}{\tau^2}+O(\tau^{-2}),~~~\tau\geq1\end{align} Utilizing the continuity condition of the aforementioned solution at $\tau=1,$ we
  obtain \begin{equation}C_1=\frac{60.30}{M^2},~~~C_2=\frac{1252.35}{M^2}
 \end{equation} for which that we can write \begin{align}&A_1(\tau)\sim- \frac{0.01}{{M}^{2}}+ \frac{44.38}{M^2}(\tau-1),~~~\tau<1\notag\\&
 ~~~~~~\sim- \frac{19.19}{M^2}\ln(2\tau),~~~\tau\geq1.\end{align} To find the explicit form of the perturbation contribution of the metric field $B_1(r)$ in limits $r\to0,$ we
 need
to obtain the asymptotic behavior of the functions $\Sigma_{1,2}(r)$ at central regions of the space time, such that \begin{align}\lim_{r\to0}\left(%
\begin{array}{c}
  \Sigma_1 \\
  \Sigma_2 \\
\end{array}%
\right)\sim\left(%
\begin{array}{c}
  -12/r^2 \\
  -75/2r^3 \\
\end{array}%
\right)\end{align} for which the integral equation (\ref{lnB1})
reads \begin{equation}\ln
B_1=\frac{75}{24}\int\frac{dr}{r}=\frac{25}{12}\int
\frac{d\tau}{\tau}=\frac{25}{12}\ln\bigg(\frac{3r^{3/2}}{Q\sqrt{M}}\bigg).\end{equation}
Also the equation (\ref{329}) has other finite non-vanishing
singular points $R_{\pm}$ and $R_{0H}$ for which the functions
(\ref{XYZ}) must satisfy the following conditions, as well.
\begin{align}&
4\Lambda R_{\pm}^4+3Q^2-4R_{\pm}^2=0\notag\\& \Lambda
R_{0H}^4+6MR_{0H}-3Q^2-3R_{0H}^2=0\end{align} where $R_{0H}$ is in
fact the horizon radius of a RN dS/AdS black hole without the
interaction part $\alpha=0$ but
\begin{equation}R_{\pm}=\sqrt{\frac{1\pm\sqrt{1-3\Lambda
Q^2}}{2\Lambda}}\end{equation} which for small cosmological
parameter  $Q^2\Lambda<<\frac{1}{3}$ reach
\begin{align}R_+\sim\sqrt{\frac{1}{\Lambda}},~~~R_-\sim Q\frac{\sqrt{3}}{2}.\end{align} Thus in order to solve the complicated second order linear homogeneous differential
equation (\ref{329}) we use dominant terms of the functions
$X,Y,Z$ by substituting
 \begin{align}\label{XYZlim}&\left(%
\begin{array}{c}
  X^{\pm}(r) \\
  Y^{\pm}(r) \\
  Z^{\pm}(r) \\
\end{array}%
\right)=\lim_{r\to R_{\pm}}\left(%
\begin{array}{c}
  X(r) \\
  Y(r) \\
  Z(r) \\
\end{array}%
\right)\sim\frac{1}{(r^2-R_{+}^{2})(r^2-R_{-}^{2})}\left(%
\begin{array}{c}
  W^{\pm} \\
  F_{\pm1}^*/2 \\
  {Q^2F_{\pm2}^*}/{R_{\pm}^3} \\
\end{array}%
\right)\notag\\&
\left(%
\begin{array}{c}
  Y_{0H}(r) \\
  Z_{0H}(r) \\
\end{array}%
\right)=\lim_{r\to
R_{0H}}\left(%
\begin{array}{c}
  Y(r) \\
  Z(r) \\
\end{array}%
\right)\sim\frac{1}{(r-R_{0H})^2}\left(%
\begin{array}{c}
  F_{0H1}^*/2 \\
  Q^2F_{0H2}^*/R_{0H}^3 \\
\end{array}%
\right)\end{align}
  in which  \begin{align}&W^{\pm}=\frac{3Q^2-2R_{\pm}^{2}}{R_{\pm}  \Lambda}\notag\\&F_{\pm1,2}^{*}=\frac{F_{1,2}(R_{\pm})}{24R^5_{\pm}\Lambda
(\Lambda\,{R_{\pm}}^{4}+6\,M
R_{\pm}-3\,{Q}^{2}-3\,{R_{\pm}}^{2})}\notag\\&F_{0H1,2}^*=\frac{F_{0H1,2}(R_{0H})}{24R_{0H}^5(2\Lambda
R^3_{0H}+3M-3R_{0H})^2(4\Lambda
R_{0H}^4+3Q^2-4R_{0H}^2)}\end{align} and we used the leading order
Taylor series expansion
\begin{align}\Lambda\,{r}^{4}+6\,Mr-3\,{Q}^{2}-3\,{r}^{2}\sim2 (2\Lambda R^3_{0H}+3M-3R_{0H})(r-R_{0H}).\end{align}
Regarding the asymptotic relations (\ref{XYZlim}) one can show
that the equation (3.29) reduces to the following form
\begin{equation}A_1^{\prime\prime}+\frac{a_{\pm}A^\prime_1}{r-R_{\pm}}+\frac{b_{\pm}A_1}{r-R_{\pm}}=\frac{c_{\pm}}{r-R_{\pm}}\end{equation} with general solution
\begin{equation}A_1=\frac{c_{\pm}}{b_{\pm}}+\frac{C_3J_{1-a_{\pm}}(2\sqrt{b_{\pm}(r-R_{\pm})})}{(r-R_{\pm})^\frac{a_{\pm}-1}{2}}+\frac{C_4N_{1-a_{\pm}}(2\sqrt{b_{\pm}(r-R_{\pm})})}{(r-R_{\pm}
)^\frac{a_{\pm}-1}{2}}\end{equation} where
\begin{align}a_{\pm}=\frac{\pm W_{\pm} }{2R_{\pm}(R_+^2-R_-^2)},~~~b_{\pm}=\frac{\pm F_{\pm1}^*}{4R_{\pm}(R_+^2-R_-^2)},~~~c_{\pm}=\frac{\pm Q^2F_{\pm2}^*}{
2R^4_{\pm}(R_+^2-R_-^2)}.\end{align} In the above solutions
$C_{3,4}$ are constants of integration and $J,N$ are the Bessel
and the Newmann functions of order $1-a$. To find solutions near
the singular point $r_{0H}$ we have
\begin{equation}A_1^{\prime\prime}+a_{0H}A^\prime_1+\frac{b_{0H}A_1}{(r-R_{0H})^2}=\frac{c_{0H}}{(r-R_{0H})^2}\end{equation}
with special solutions
\begin{align}\label{A11}A_1(r)&={\frac
{c_{0H}}{b_{0H}}}+C_5(r-R_{0H})^{\epsilon_+}+C_6(r-R_{0H})^{\epsilon_-},
\end{align}
in which
\begin{align}&
\epsilon_{\pm}=\frac{(1-a_{0H})}{2}\bigg[1\mp\sqrt{1-\frac{4b_{0H}}{(1-a_{0H})^2}}\bigg],\notag\\&
a_{0H}=
X(R_{0H}),~~~b_{0H}=\frac{F_{0H1}^*}{2},~~~c_{0H}=\frac{Q^2F_{0H2}^*}{R_{0H}^3}.\end{align}
To determine $C_{5,6}$ in (\ref{A11}) we apply the continuity
condition of the solutions on the particular hyper-surfaces
$R_{\pm}$ and $R_{0H}.$ This is done trivially same as one which
we found for former case (calculated near the hyper-surface
$\tau=1$). Thus we will not proceed to find the coefficients
$C_{5,6}$ now but we calculate equation of state of modified RN
AdS black hole metric for solution (\ref{337}) in the next
section.
\section{Equation of state}
Before finding the equation of state of the black hole under
consideration we collect our obtained perturbation solutions of
the fields in the previous section, such that
\begin{align}&E(r)\sim E_0(r)+\alpha E_1(r)=\frac{Q(1-4\alpha\Lambda)}{r^2}+\frac{4\alpha Q^3}{r^6}\end{align}
\begin{align}&A(r)\sim A_0(r)+\alpha
A_1(r),~~~r<\bigg(\frac{Q\sqrt{M}}{3}\bigg)^\frac{2}{3}\notag\\&~~~~~~~=1-\frac{2M}{r}+\frac{Q^2}{r^2}-\frac{\Lambda
r^2}{3}-\frac{44.38\alpha}{M^2}\bigg(1-\frac{3r^{3/2}}{Q\sqrt{M}}\bigg)
\end{align}\begin{align}&A(r)\sim A_0(r)+\alpha
A_1(r),~~~r\geq\bigg(\frac{Q\sqrt{M}}{3}\bigg)^\frac{2}{3}\notag\\&~~~~~~~=1-\frac{2M}{r}+\frac{Q^2}{r^2}-\frac{\Lambda
r^2}{3}-\frac{19.19\alpha}{M^2}\ln\bigg(\frac{3r^{3/2}}{Q\sqrt{M}}\bigg)
\end{align}and \begin{align}
B(r)\sim
B_{0}(r)+B_{1}(r)=\bigg(1-\frac{2M}{r}+\frac{Q^2}{r^2}-\frac{\Lambda
r^2}{3}\bigg)^{-1}+\alpha\bigg(\frac{3r^{3/2}}{Q\sqrt{M}}\bigg)^{\frac{25}{12}}\end{align}
for both regions  $r<\big(\frac{Q\sqrt{M}}{3}\big)^\frac{2}{3}$
and $r\geq\big(\frac{Q\sqrt{M}}{3}\big)^\frac{2}{3}.$
 To obtain the equation of state of the modified RN-AdS black
hole we must first solve the event horizon equation
$g_{tt}(r_+)=-A(r_+)=0,$ such that
\begin{align}\label{Hor}
&1-\frac{2M}{r_+}+\frac{Q^2}{r_+^2}-\frac{\Lambda
r_+^2}{3}-\frac{44.4\alpha}{M^2}\bigg(1-\frac{3r_+^{3/2}}{Q\sqrt{M}}\bigg)\cong0,~~~r_+<\bigg(\frac{Q\sqrt{M}}{3}\bigg)^\frac{2}{3}
\notag\\&1-\frac{2M}{r_+}+\frac{Q^2}{r_+^2}-\frac{\Lambda
r_+^2}{3}-\frac{20\alpha}{M^2}\ln\bigg(\frac{3r_+^{3/2}}{Q\sqrt{M}}\bigg)\cong0,~~~r_+\geq\bigg(\frac{Q\sqrt{M}}{3}\bigg)^\frac{2}{3}
\end{align}
where $r_+$ is the radius of the exterior horizon.  The Hawking
temperature of the black hole system is found vs the surface
gravity of the horizon such that
\begin{align}\label{tem}&T_{<}=\frac{1}{8\pi}\frac{d
g_{tt}}{dr}\bigg|_{r_+}\cong\frac{1}{4\pi}\bigg[\frac{M}{r_+^2}-\frac{Q^2}{r_+^3}-\frac{\Lambda
r_+}{3}+\frac{99\alpha
r_+^{1/2}}{QM^2\sqrt{M}}\bigg],~~~r_+<\bigg(\frac{Q\sqrt{M}}{3}\bigg)^\frac{2}{3}\notag\\&T_{\geq}=\frac{1}{8\pi}\frac{d
g_{tt}}{dr}\bigg|_{r_+}\cong\frac{1}{4\pi}\bigg[\frac{M}{r_+^2}-\frac{Q^2}{r_+^3}-\frac{\Lambda
r_+}{3}-\frac{15\alpha
}{M^2r_+}\bigg],~~~r_+\geq\bigg(\frac{Q\sqrt{M}}{3}\bigg)^\frac{2}{3}.
\end{align} For a quantum perturbed evaporating black hole the enthalpy which is defined by the black hole mass is not a constant.
Thus,
 when we investigate possible phase transition
then, we must eliminate $M$ in the Hawking temperature above via
the horizon equation (\ref{Hor}). But we cannot obtain a closed
form for $M$ in this way and so we must use a particular technical
method. To do so, we call (\ref{336}), such that
\begin{equation}\label{mass}M=\frac{9r_+^3}{Q^2\tau^2},~~~-\infty<\tau<\infty\end{equation} and substitute into the horizon equation
(\ref{Hor}) and the corresponding Hawking temperature (\ref{tem})
to obtain
\begin{align}\label{Hor1}
&1-\frac{18r_+^2}{Q^2\tau^2}+\frac{Q^2}{r_+^2}-\frac{\Lambda
r_+^2}{3}-\frac{44.4\alpha Q^4\tau^4}{81
r_+^6}(1-\tau)\cong0,~~~\tau<1\notag\\&1-\frac{18r_+^2}{Q^2\tau^2}+\frac{Q^2}{r_+^2}-\frac{\Lambda
r_+^2}{3}-\frac{20\alpha Q^4\tau^4}{81
r_+^6}\ln\tau\cong0,~~~\tau\geq1
\end{align} with
\begin{align}\label{tem1}&T_{<}=\frac{1}{8\pi}\frac{d
g_{tt}}{dr}\bigg|_{r_+}\cong\frac{1}{4\pi}\bigg[\frac{9r_+}{Q^2\tau^2}-\frac{Q^2}{r_+^3}-\frac{\Lambda
r_+}{3}+\frac{11\alpha Q^4\tau^5 }{27
r_+^7}\bigg],~~~\tau<1\notag\\&T_{\geq}=\frac{1}{8\pi}\frac{d
g_{tt}}{dr}\bigg|_{r_+}\cong\frac{1}{4\pi}\bigg[\frac{9r_+}{Q^2\tau^2}-\frac{Q^2}{r_+^3}-\frac{\Lambda
r_+}{3}-\frac{5\alpha Q^4\tau^4}{27r^7_+}\bigg],~~~\tau\geq1
\end{align}
in which $\tau$ plays role of an additional parameter. We are now
in the position to write the  equation of state of modified RN AdS
black hole by using the above relationships. To do so we first
define dimensionless quantities
\begin{align}\label{dim}v=\frac{r_+}{Q},~~~t=4\pi QT,~~~p=-\frac{\Lambda Q^2}{3}~~~\bar{\alpha}=\frac{11\alpha}{27Q^2}\end{align}
 in which $v$ is specific volume, $t$ is temperature and $p>0$ is AdS space pressure. Using these dimensionless thermodynamics variables the Hawking temperature
(\ref{tem1}) is simplified to the following equation of state.
\begin{align}\label{tem2}&t_{<}=pv+\frac{9v}{\tau^2}-\frac{1}{v^3}+\frac{\bar{\alpha}\tau^5
}{v^7},~~~\tau<1\notag\\&t_{\geq}=pv+
\frac{9v}{\tau^2}-\frac{1}{v^3}-\frac{5\bar{\alpha
}\tau^4}{297v^7},~~~\tau\geq1
\end{align} with corresponding horizon equations\begin{align}\label{Hor2}
&1-\frac{18v^2}{\tau^2}+\frac{1}{v^2}+pv^2-\frac{4\bar{\alpha}\tau^4(1-\tau)}{3v^6}\cong0,~~~\tau<1\notag\\&1-\frac{18v^2}{\tau^2}+\frac{1}{v^2}+pv^2-\frac{20\bar{\alpha}
\tau^4\ln\tau}{33v^6}\cong0,~~~\tau\geq1.
\end{align}
  These horizon equations serve constraint conditions for the equation of states above in both
regions and for $\tau=1,$ they reach to a similar horizon equation
and the Hawking temperature. Consequently,  we study the
possibility of phase transition of the modified RN AdS black hole
for both cases $\tau<1$ and $\tau\geq1$ separately in what
follows. To do so, we first eliminate $\bar{\alpha}$ in the
Hawking temperature by the horizon equation, such that
\begin{align}\label{state2}&t_{\tau<1}=\bigg(\frac{\tau-4}{\tau-1}\bigg)\frac{pv}{4}+\frac{3\tau}{4(1-\tau)}\frac{1}{v}+\frac{1}{4}
\bigg(\frac{4-7\tau}{\tau-1}\bigg)\frac{1}{v^3}+\bigg(\frac{5\tau-2}{\tau-1
}\bigg)\frac{9v}{2}\notag\\&
\bar{\alpha}=\frac{3pv^8}{4\tau^4(1-\tau)}- \frac{3v^4(\tau^2
v^2-18v^4+\tau^2)}{4\tau^6(\tau-1)}\end{align} and
\begin{align}\label{state3}&t_{\tau\geq1}=\bigg(1-\frac{1}{36\ln\tau}\bigg)pv+\bigg(9+\frac{1}{2\ln\tau
}\bigg)\frac{v}{\tau^2}-\frac{1}{36\ln\tau
v}-\bigg(1+\frac{1}{36\ln\tau}\bigg)\frac{1}{v^3}\notag\\&
\bar{\alpha}=\frac{33v^4(pv^4\tau^2+v^2\tau^2-18v^4+\tau^2)}{20\tau^6\ln\tau}\end{align}
\section{Phase transition}
Here, we are interested into study the existence of multiple
phases and the possible phase transitions between them. Therefore,
it is first necessary to determine critical points in the phase
space that indicate the possibility of different phases and/or
even coexistence between them for the black hole system. The
critical points in three dimensional phase space $(p,v,t)$ at
constant pressure are obtained by solving the critical equations
\begin{equation}\label{CE}
    \left(\frac{\partial t}{\partial v} \right) _{p}=0,~~~~ \left(\frac{\partial^2 t}{\partial v^2} \right)
    _{p}=0.
\end{equation} In order to study the phase transition of the modified RN-AdS
black hole, we employ the equation of states (\ref{tem2}) in two
different cases $\tau<1$ and $\tau>1$ respectively. The following
is addressed a detailed description of the methodology employed.
\subsection{Phase transition in case of $|\tau|<1$} Substituting $t_<$
given by (\ref{state2}) and solving the equations of critical
point (\ref{CE}) we find parametric critical points as
\begin{align}\label{crit}
&v_c^\pm=\pm\sqrt{2\bigg(\frac{4}{\tau}-7\bigg)}
,p^\pm_c=\frac{3(\tau^4-840\tau^2+816\tau-192)}{4\tau^2(\tau-4)(7\tau-4)
},\notag\\&t^\pm_c=\frac{\pm\tau}{1-\tau}\sqrt{\frac{\tau}{2(7\tau-4)}}
\end{align}
where negative (positive) sign corresponds to negative electric
charge $Q<0(>0).$ To find permissable numeric values for
$\bar{\alpha}$ in case of $|\tau|<1,$ we must substitute the
parametric critical points above into the horizon equation given
by (\ref{Hor2}) such that
\begin{equation}\label{alphac}\bar{\alpha}^\pm_c=-24\,{\frac { \left( 7\,\tau-4 \right) ^{2} \left( {\tau}^{5}+6\,{\tau
}^{4}-2650\,{\tau}^{3}+5670\,{\tau}^{2}-3888\,\tau+864 \right)
}{{\tau }^{10} \left( \tau-4 \right)  \left( -1+\tau \right) }}.
\end{equation}
We plotted diagrams of the critical parameters above vs $\tau$ in
figure 1.
\begin{figure} \centering
 \subfigure[{}]{\label{figalpahc}
 \includegraphics[width=0.32\textwidth]{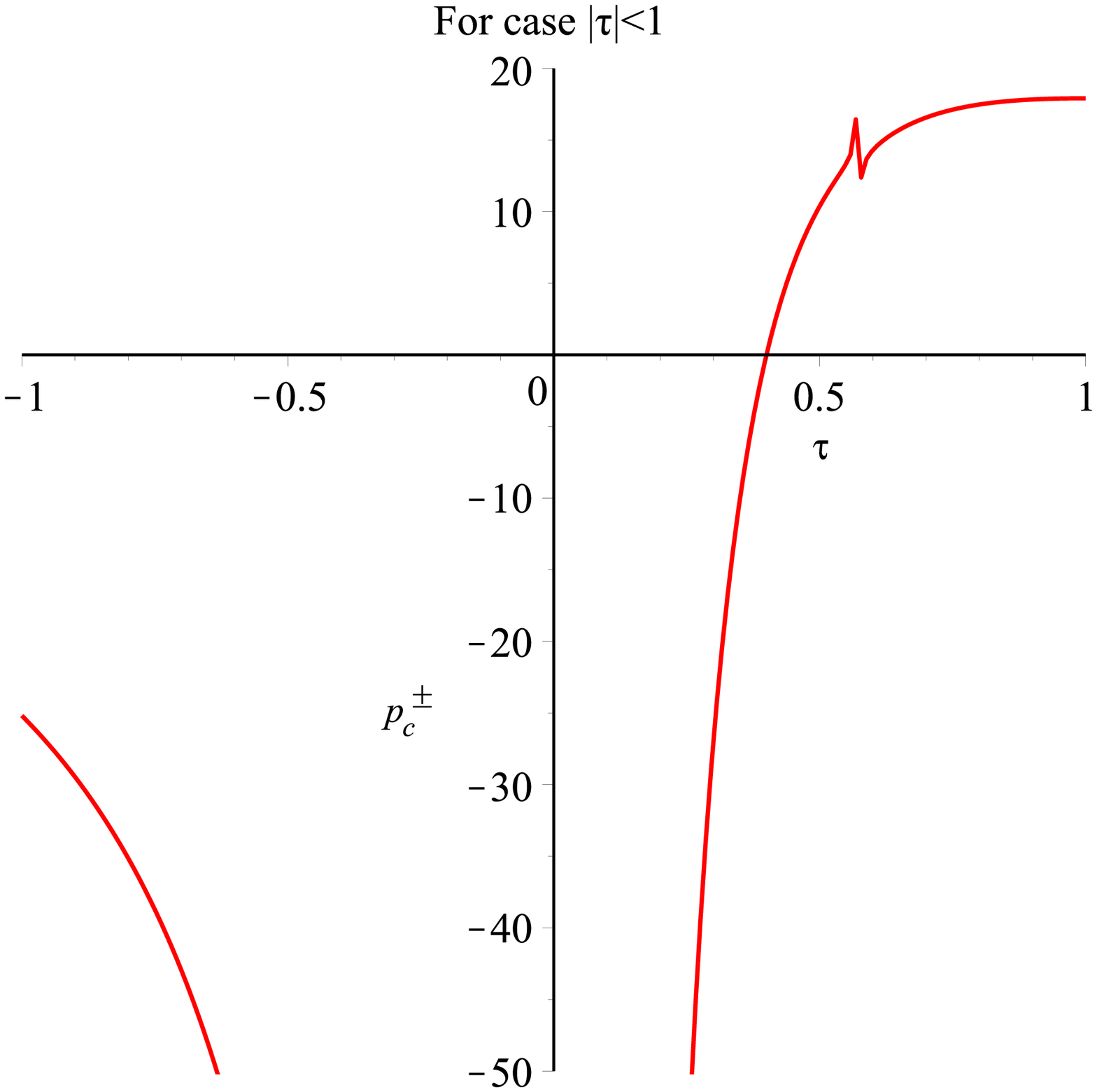}}
 \hspace{1mm}\subfigure[{}]{\label{figvc}
 \includegraphics[width=0.32\textwidth]{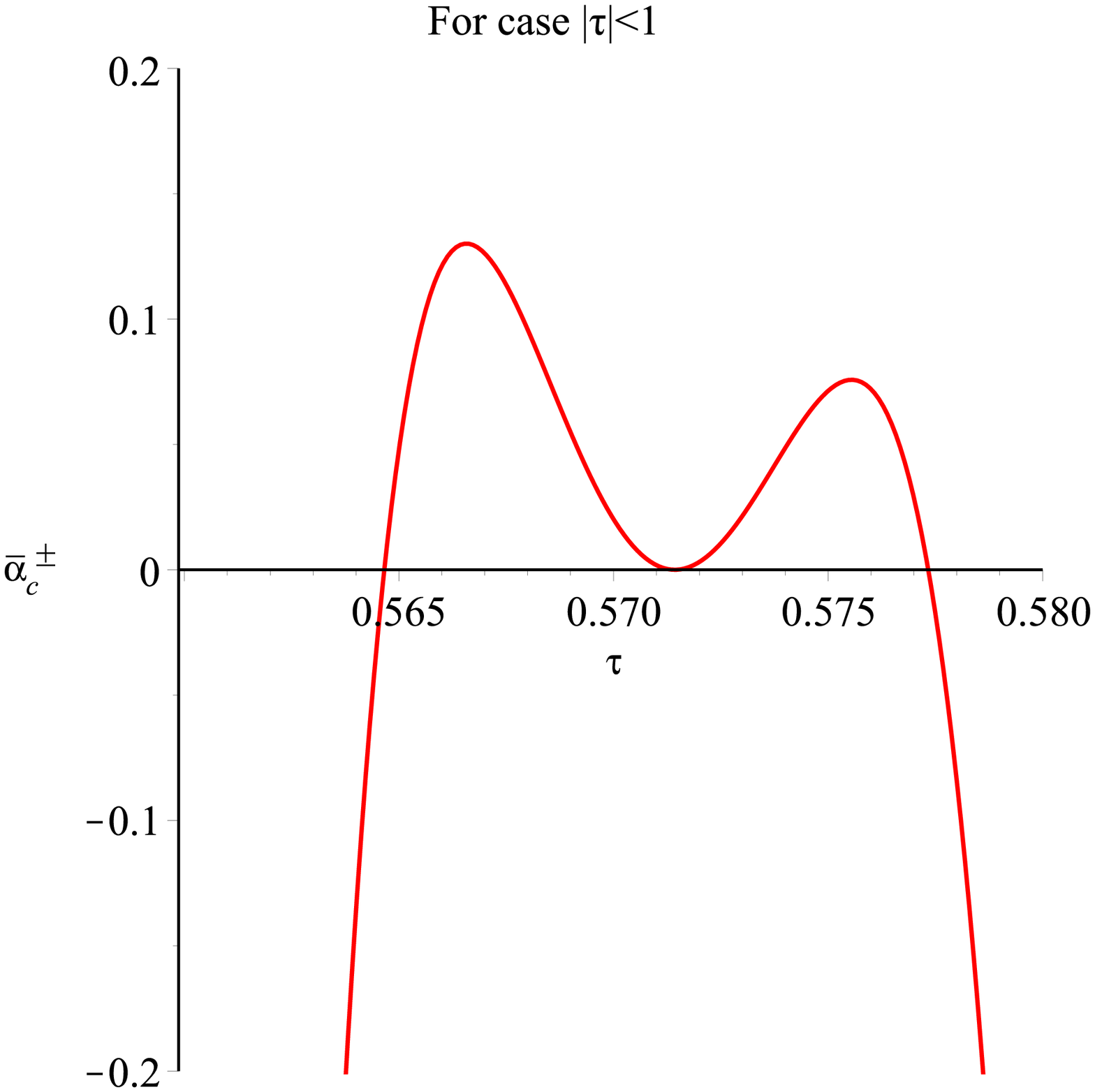}}
 \hspace{1mm}\subfigure[{}]{\label{fig11pc}
 \includegraphics[width=0.32\textwidth]{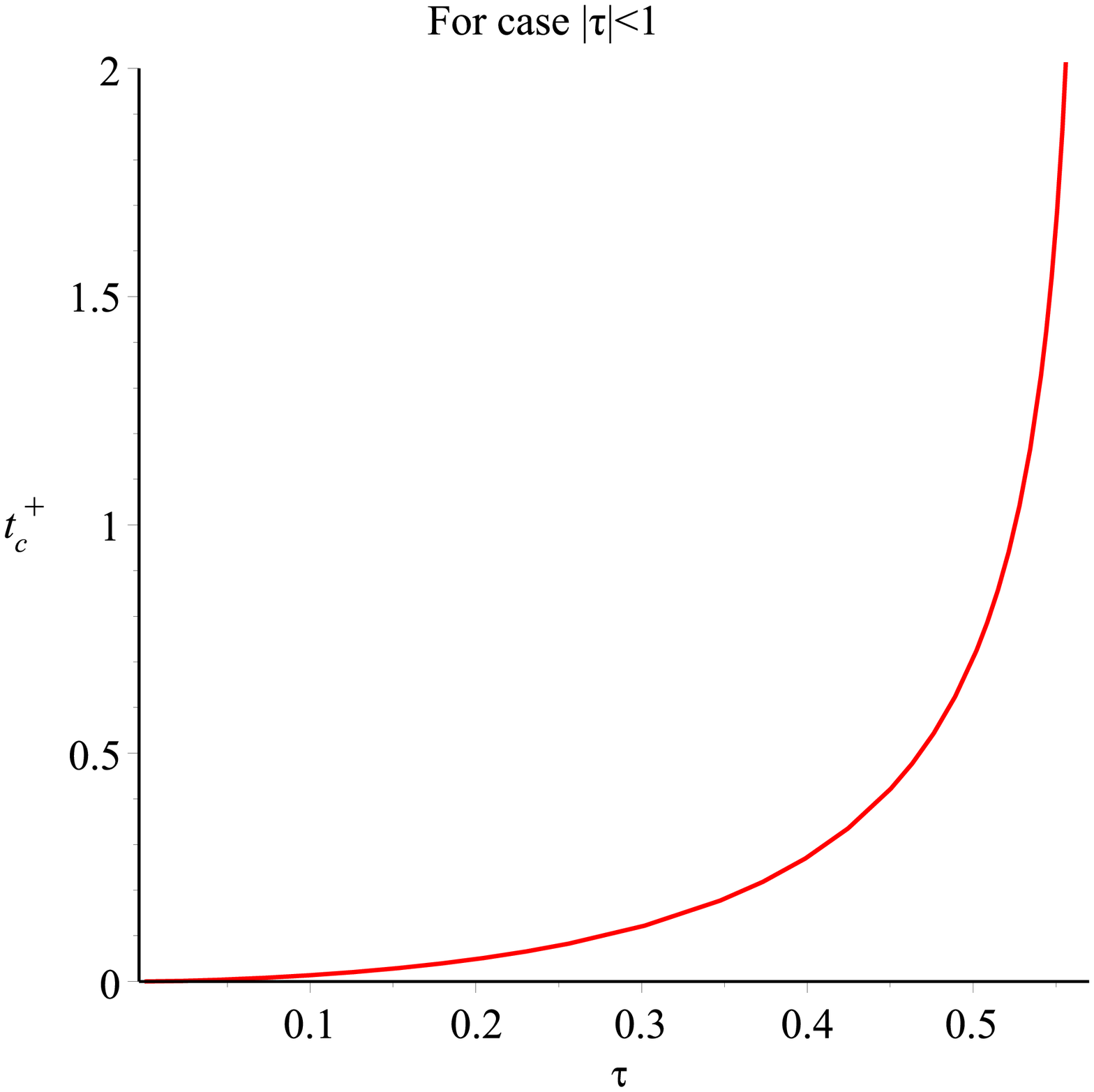}}
 \hspace{1mm}\subfigure[{}]{\label{fig12tc}
 \includegraphics[width=0.32\textwidth]{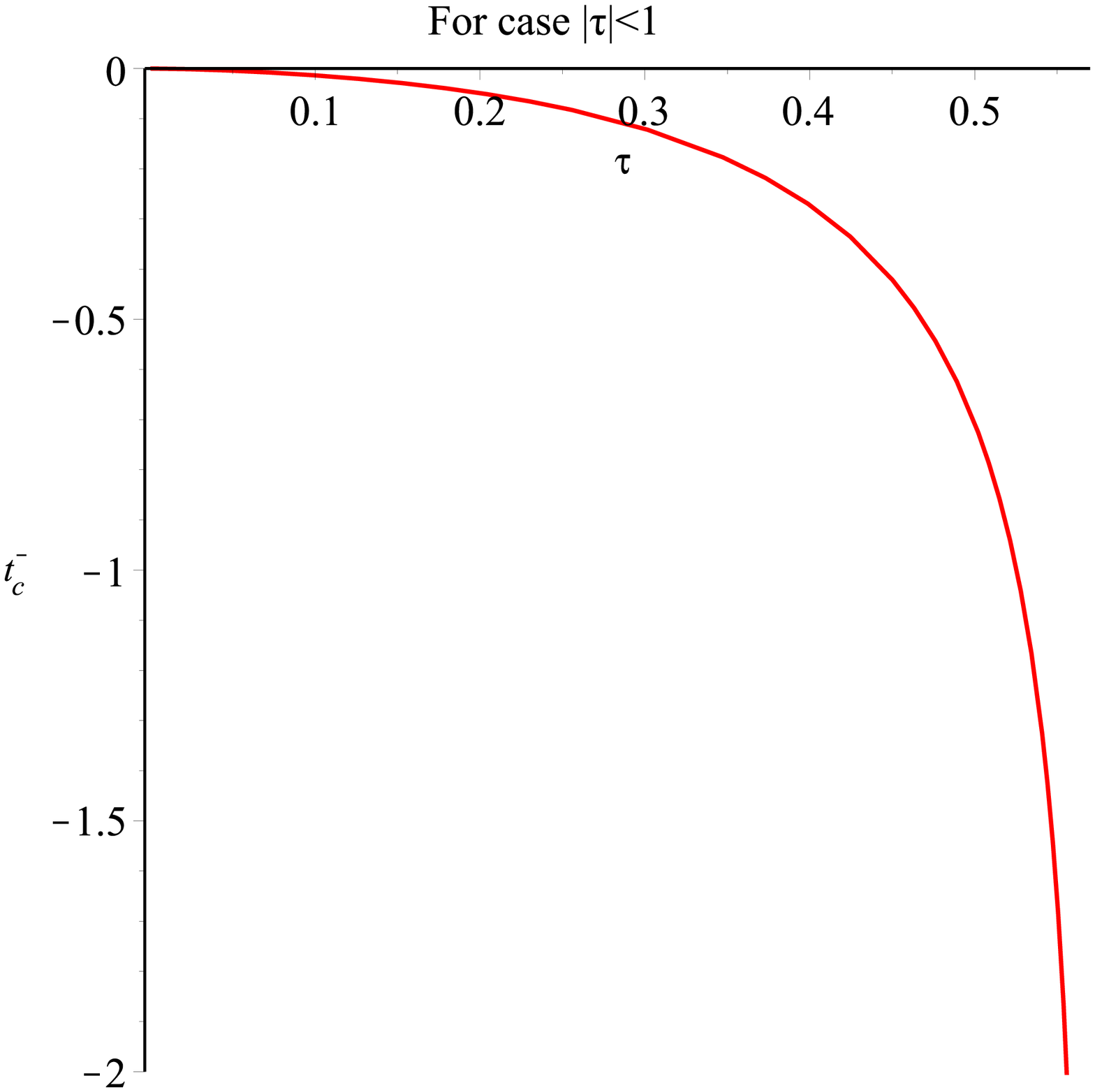}}
 \hspace{1mm}\subfigure[{}]{\label{fig13pc}
 \includegraphics[width=0.32\textwidth]{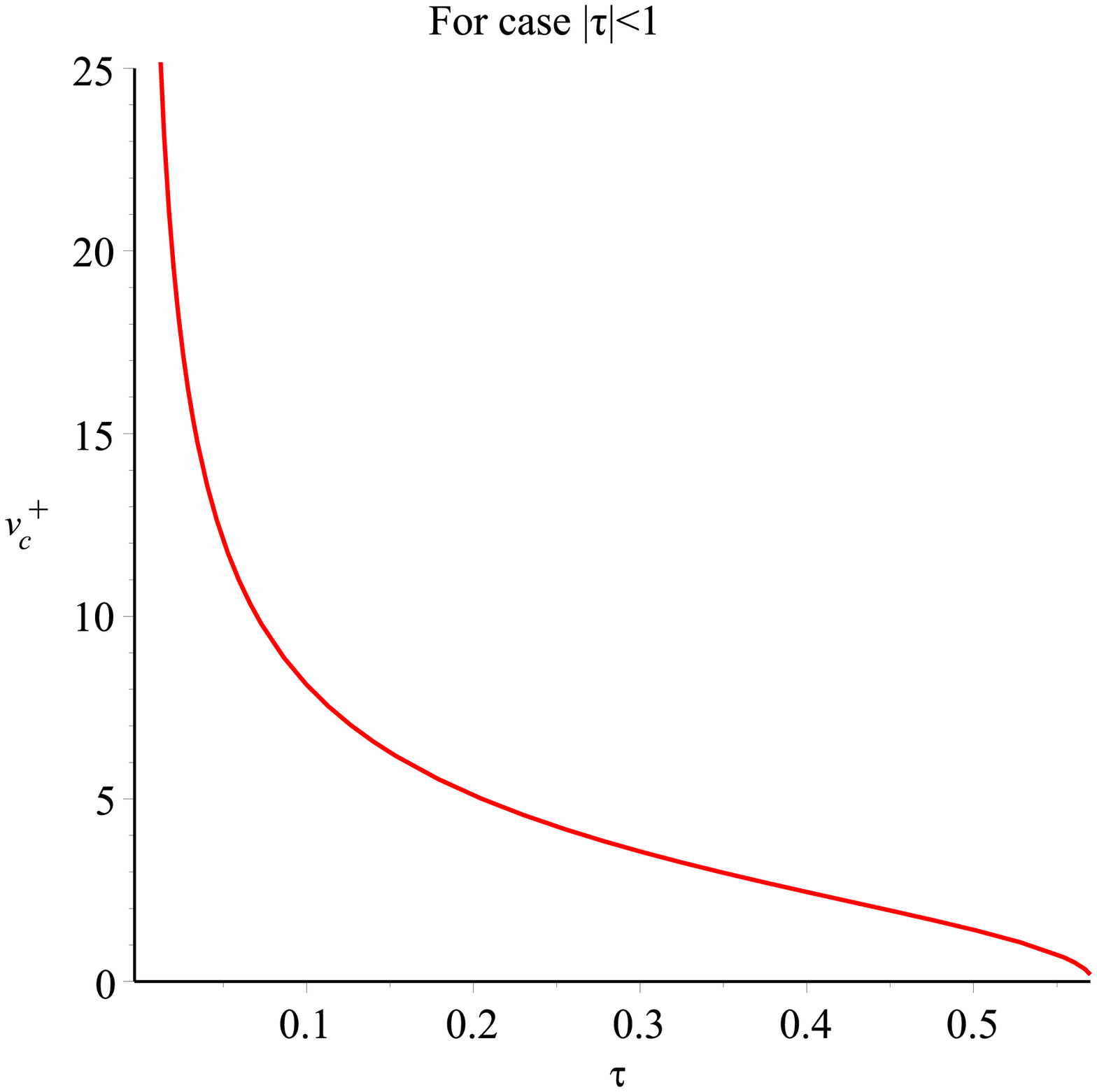}}
 \hspace{1mm}\subfigure[{}]{\label{fig14tc}
 \includegraphics[width=0.32\textwidth]{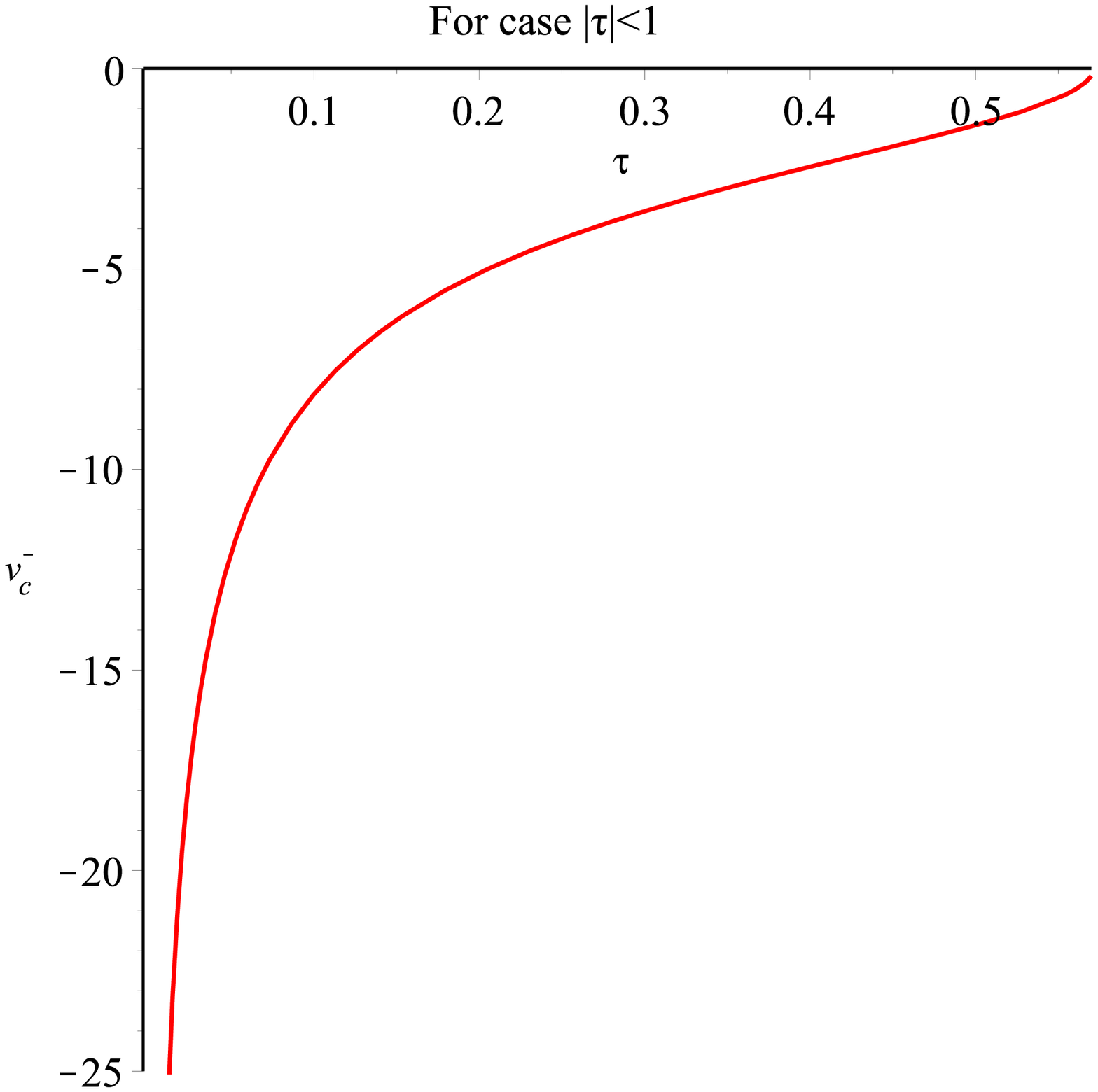}}
 \hspace{1mm}
 \caption{The critical points and the interaction parameter are plotted vs $\tau$ for $|\tau|<1$.}
\end{figure}
Looking at the figure 1 one infer that the diagrams for $Q<0$
(negative sign) are identical to those for $Q>0$ (positive sign).
consequently,  we show phase diagrams just for $Q>0$. In this
case, the real numerical values of $\bar{\alpha}$ or $\tau$ are
noted that can be chosen from the diagram displayed in Figure 1-b.
The numerical values presented in Table 1 are employed for the
plotting of phase diagrams, as illustrated in Figure 2.
\begin{center}
Table 1: Numerical values of the critical points for $\tau=0.5$
\begin{tabular}{|c|c|c|c|c|}
  \hline
  $\tau$ & $\bar{\alpha}_c$ & $t_c$ & $v_c$ & $p_c$  \\
  \hline
  0.5 & -23369.14281 & 0.7071068100 & 1.414213562 & 10.39285714 \\
  \hline
\end{tabular}
\end{center}
\begin{figure}\centering
 \subfigure[{}]{\label{figalpahc1}
 \includegraphics[width=0.4\textwidth]{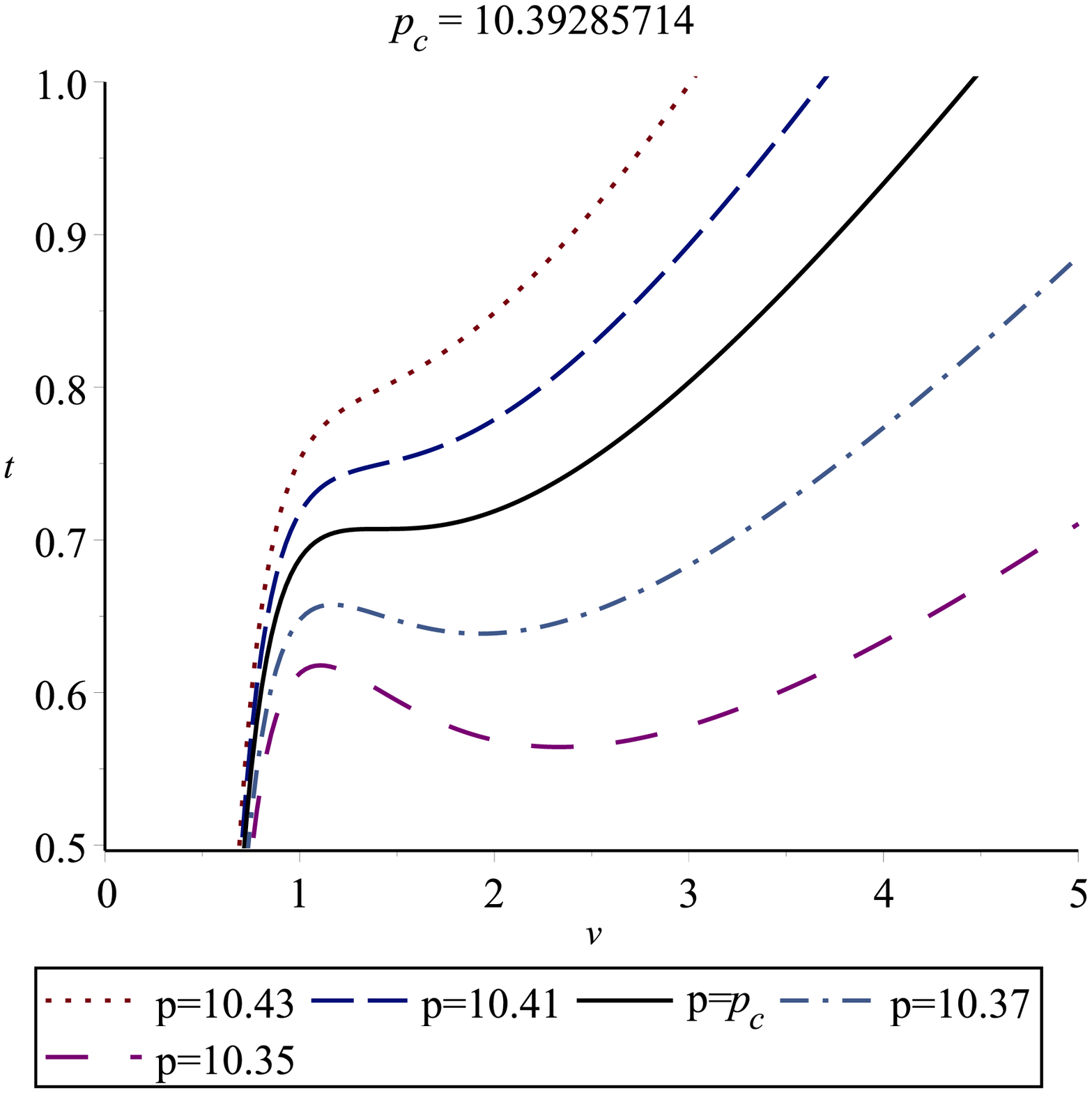}}
 \hspace{2mm}\subfigure[{}]{\label{figvc1}
 \includegraphics[width=0.4\textwidth]{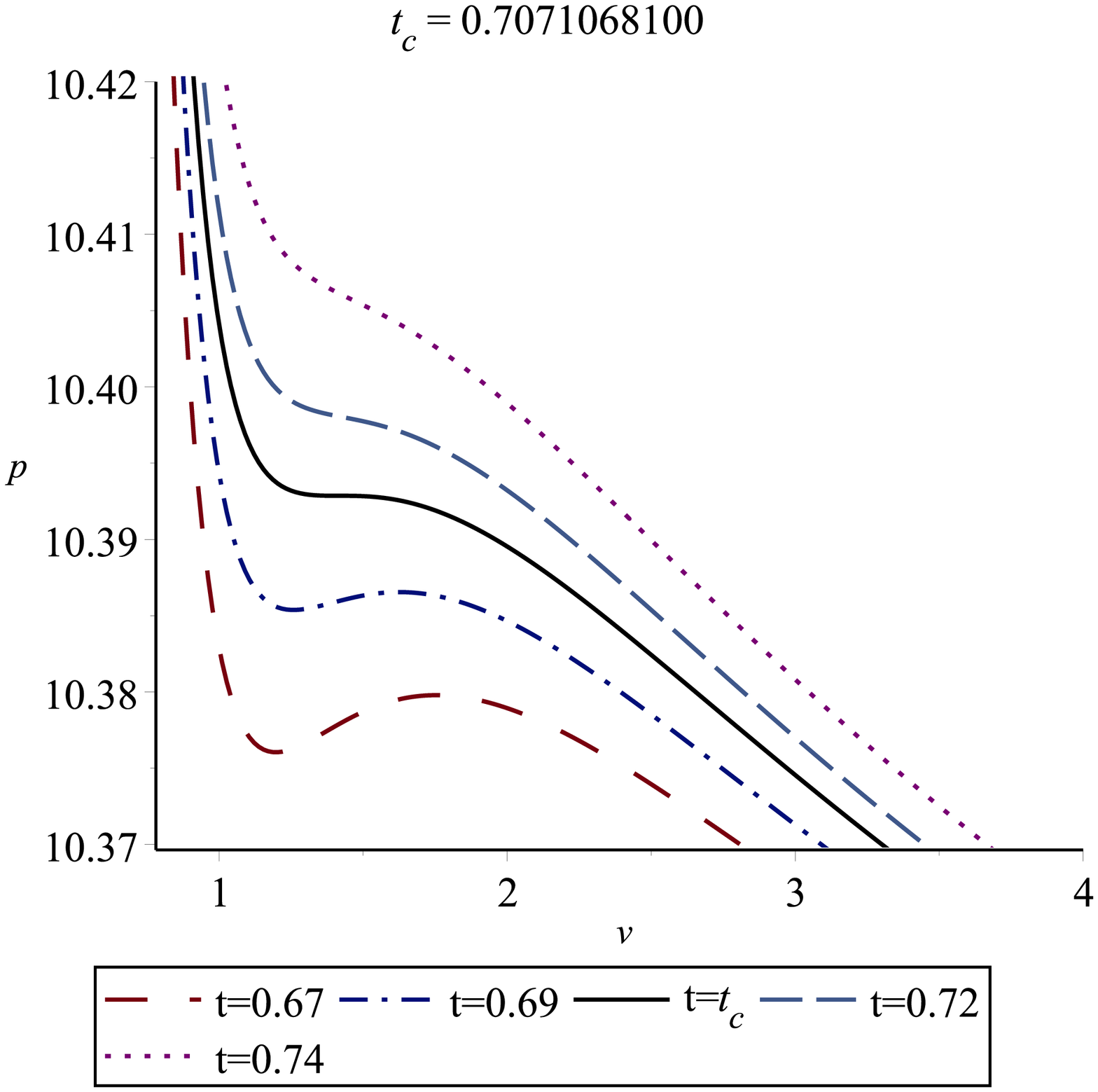}}
 \hspace{2mm}\caption{\footnotesize{ (a) t-v diagram at constant pressure and (b) p-v diagram at constant temperature are plotted for case $|\tau|<1$.}}
 \end{figure}
 We should point that in the Table 1, negative (positive) values for the critical specific
volume $v_c$ and
    the critical temperatures $t_c$ correspond with negative (positive) electric charge. The diagrams of t-v and p-v at constant pressure and constant temperature respectively
    given in the figure 2, show a small to large black hole phase
    transition which physically can be interpreted same as the
   Van der
    Waals gas-fluid phase transition in an ordinary thermodynamic
    system. There are three phases between the
    extremum points of the curves (solid/liquid/gas) at below the critical points $t<t_c$ and $p<p_c$ and for above of the critical points, the system behaves as an ideal
    gas. We emphasize that this phase transition is a small to large black hole, not the other way around. To understand this argument, we  plot
     the changes in the Gibbs energy of the system vs the temperature and specific volume, as shown in Figure 3.
Because negative values of the Gibbs free energy show
thermodynamic stability of the system but for regions with
positive Gibbs free energies, the system will be unstable. The
Gibbs free energy is defined by the equation $G=M-TS$ in which $M$
is black hole mass or alternatively, the enthalpy and $T$ is the
temperature and $S$ is the entropy. In this way, we use the
Bekenstein entropy for the black hole in absence of the Hawking
logarithmic corrections in which the entropy is equal to a quarter
of the surface area of the black hole horizon in the geometric
units. The Bekenstein entropy in the dimensionless form can be
written as follows.
\begin{equation}\label{enth}
    s=\pi v^2,~~~ s=\frac{S_B}{Q^2}.
\end{equation}
The dimensionless form of mass is obtained from \eqref{mass}, such
that
\begin{equation}
    m=\frac{9v^3}{\tau^2},~~~ m=\frac{M}{Q}
\end{equation}
Using the above relations and the temperature (\ref{state2}), we
obtain the dimensionless form for the Gibbs free energy by
$g=m-ts$ for the modified RN-AdS black holes, which in the case of
$|\tau|<1$ is
\begin{align}\label{gibss}
    &g_{\tau<1}=\bigg[\frac{9}{\tau^2}-\frac{9\pi}{2}\bigg(\frac{5\tau-2}{\tau-1}
    \bigg)-\frac{\pi p}{4}\bigg(\frac{\tau-4}{\tau-1}\bigg)
    \bigg]v^3-\frac{3\pi\tau}{4(1-\tau)}v-\frac{\pi}{4}\bigg(\frac{4-7\tau}{\tau-1}\bigg)\frac{1}{v}\end{align}
We plotted diagrams of g-t at constant pressure and g-v at
constant temperature in figure 3, for numerical critical value
given in the Table 1.
\begin{figure}\centering
\subfigure[{}]{\label{fig1pc1}
 \includegraphics[width=0.4\textwidth]{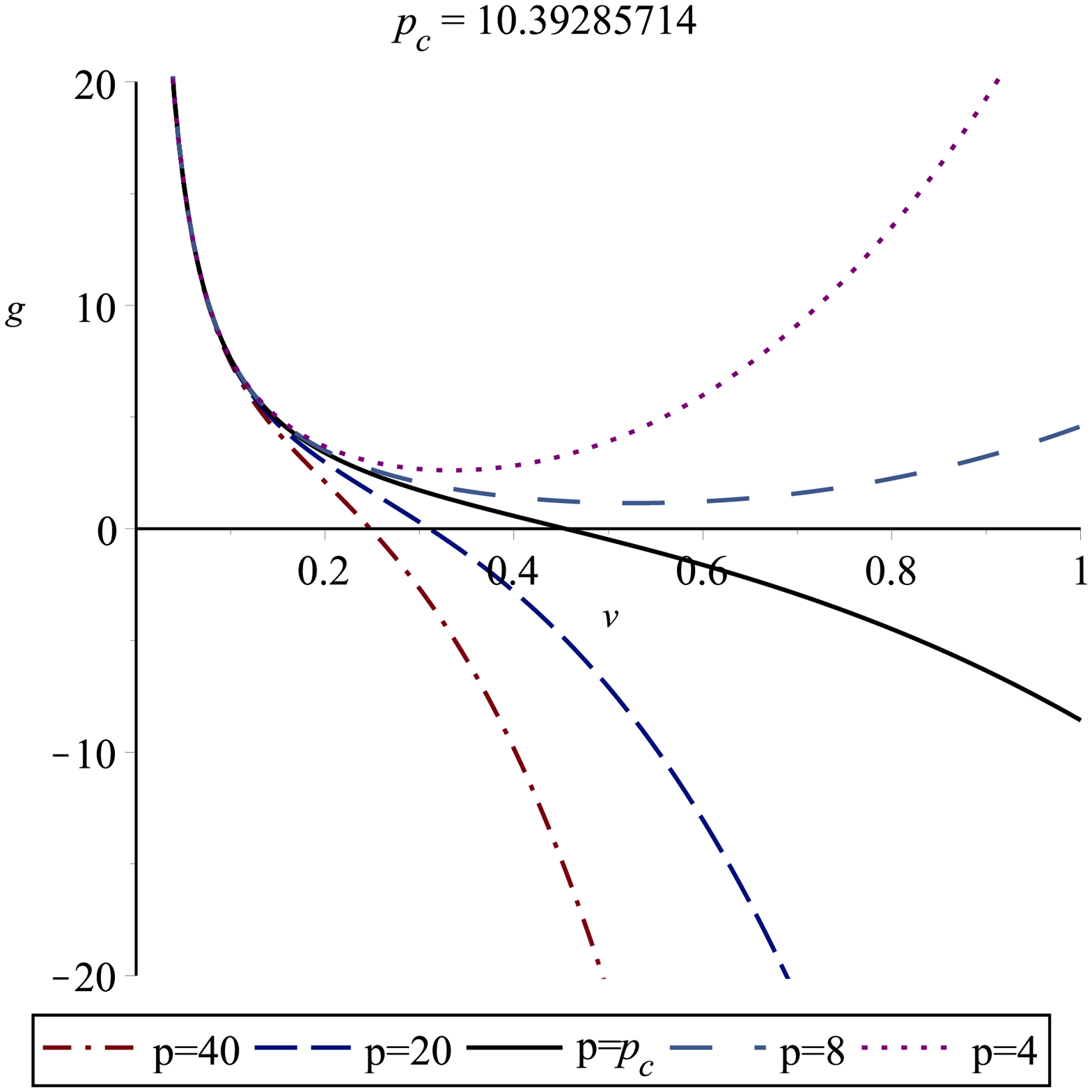}}
 \hspace{2mm}\subfigure[{}]{\label{fig1tc1}
 \includegraphics[width=0.4\textwidth]{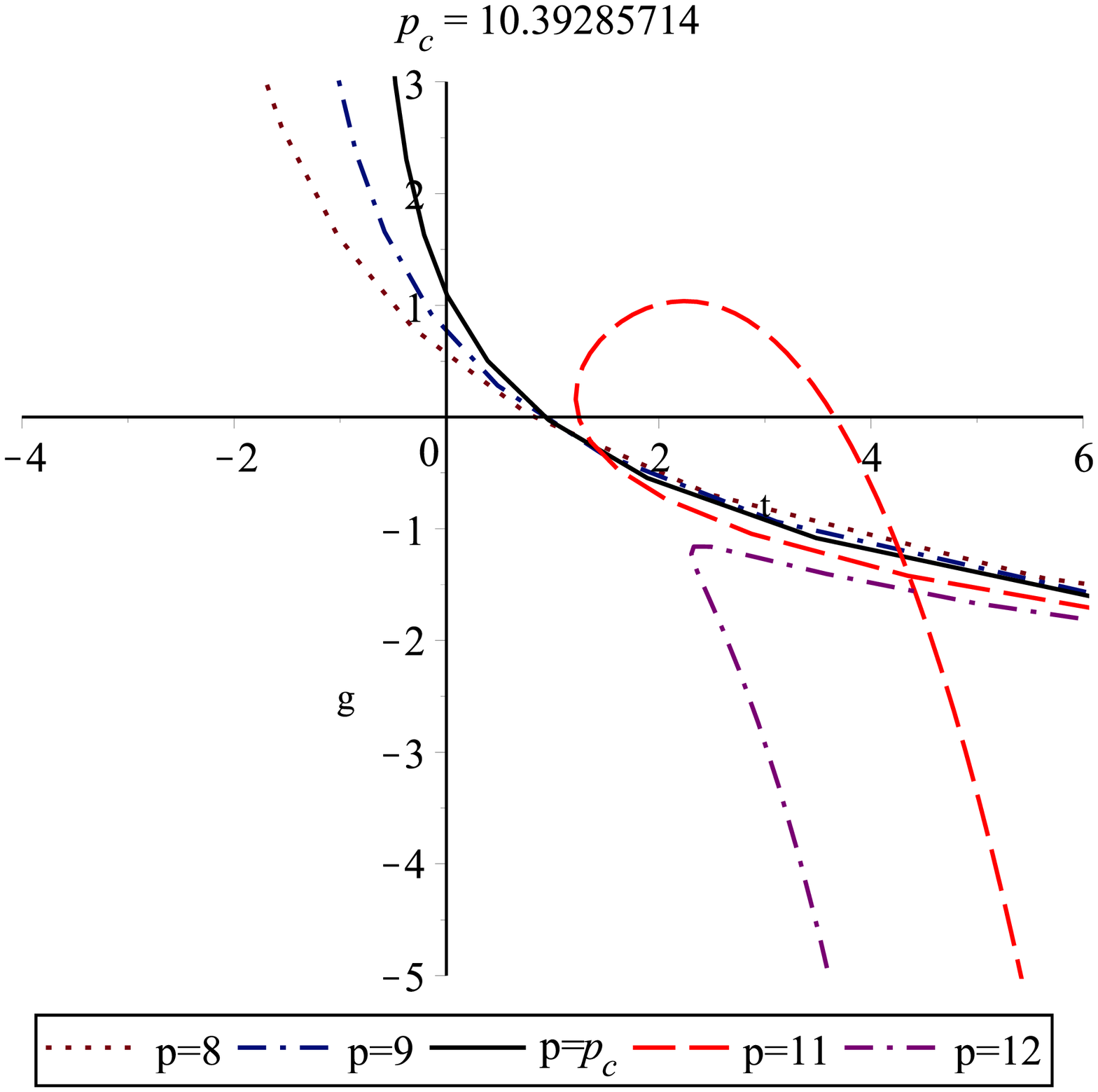}}
 \hspace{3mm}\subfigure[{}]{\label{figpc1}}
 \caption{\footnotesize{ (a) g-v and (b) g-t curves at constant pressures given in the Table 1}}
 \end{figure}
Other thermodynamic quantity which is useful to study
thermodynamic behavior of an evaporating black hole is the change
of heat capacity at constant AdS pressure which in the
dimensionless form is defined by $c_p=t\big(\frac{\partial
s}{\partial t}\big)_p. $ This is obtained by substituting
(\ref{state2}) and the entropy (\ref{enth}), such that
\begin{align}&\frac{c_p}{2\pi}=\frac{\big(\frac{\tau-4}{\tau-1}\big)\frac{pv}{4}+\frac{3\tau}{4(1-\tau)}\frac{1}{v}+\frac{1}{4}
\big(\frac{4-7\tau}{\tau-1}\big)\frac{1}{v^3}+\big(\frac{5\tau-2}{\tau-1
}\big)\frac{9v}{2}}{\big(\frac{\tau-4}{\tau-1}\big)\frac{p}{4}-\frac{3\tau}{4(1-\tau)}\frac{1}{v^2}-\frac{3}{4}\big(\frac{4-7\tau}{\tau-1}\big)\frac{1}{v^4}+\frac{9}{2}\big(\frac{5\tau-2}{\tau-1}\big)}\end{align}
\begin{figure}\centering
\subfigure[{}]{\label{fig11pc1}
  \includegraphics[width=0.4\textwidth]{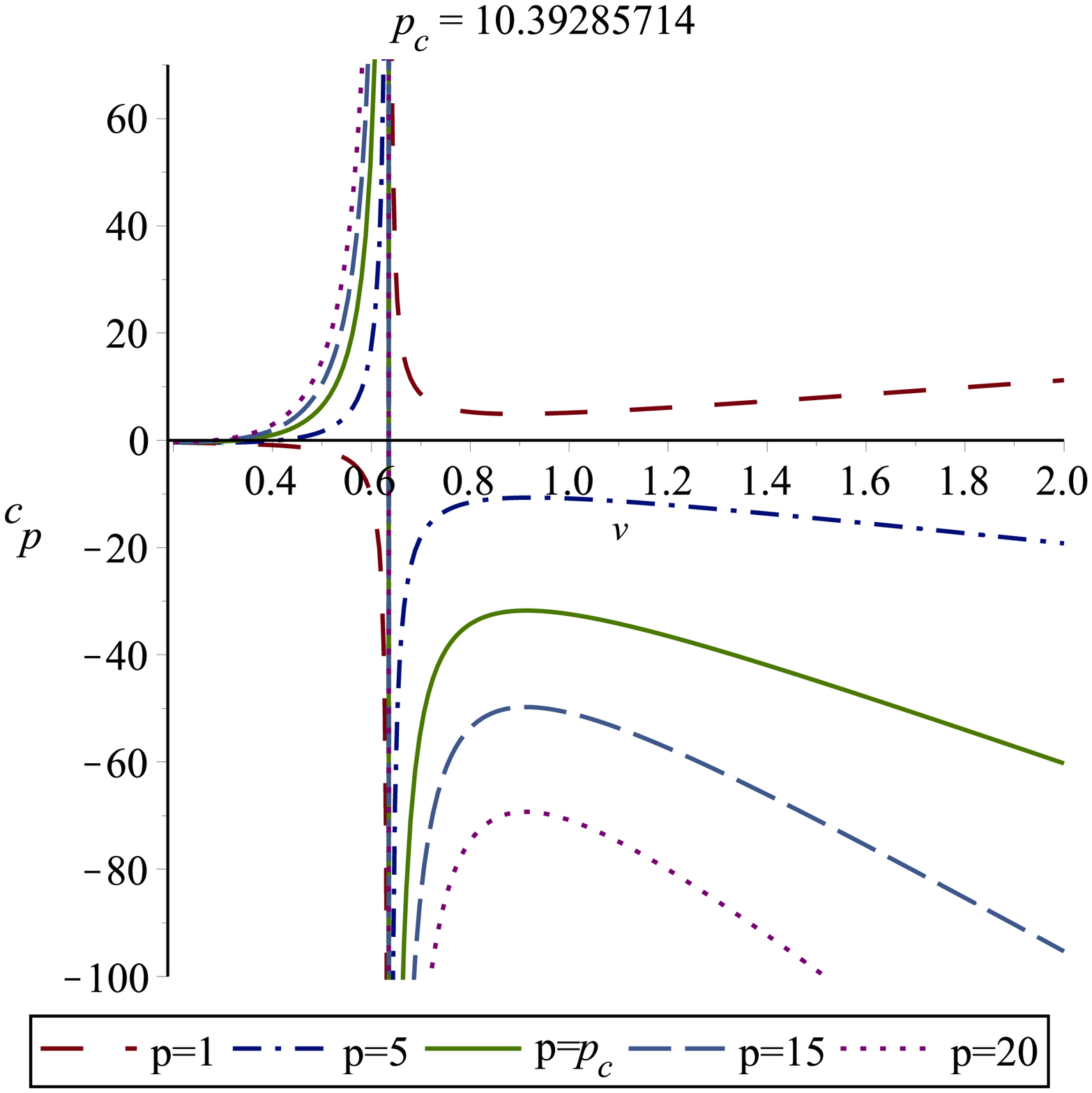}}
 \hspace{2mm}\subfigure[{}]{\label{fig11tc1}
 \includegraphics[width=0.4\textwidth]{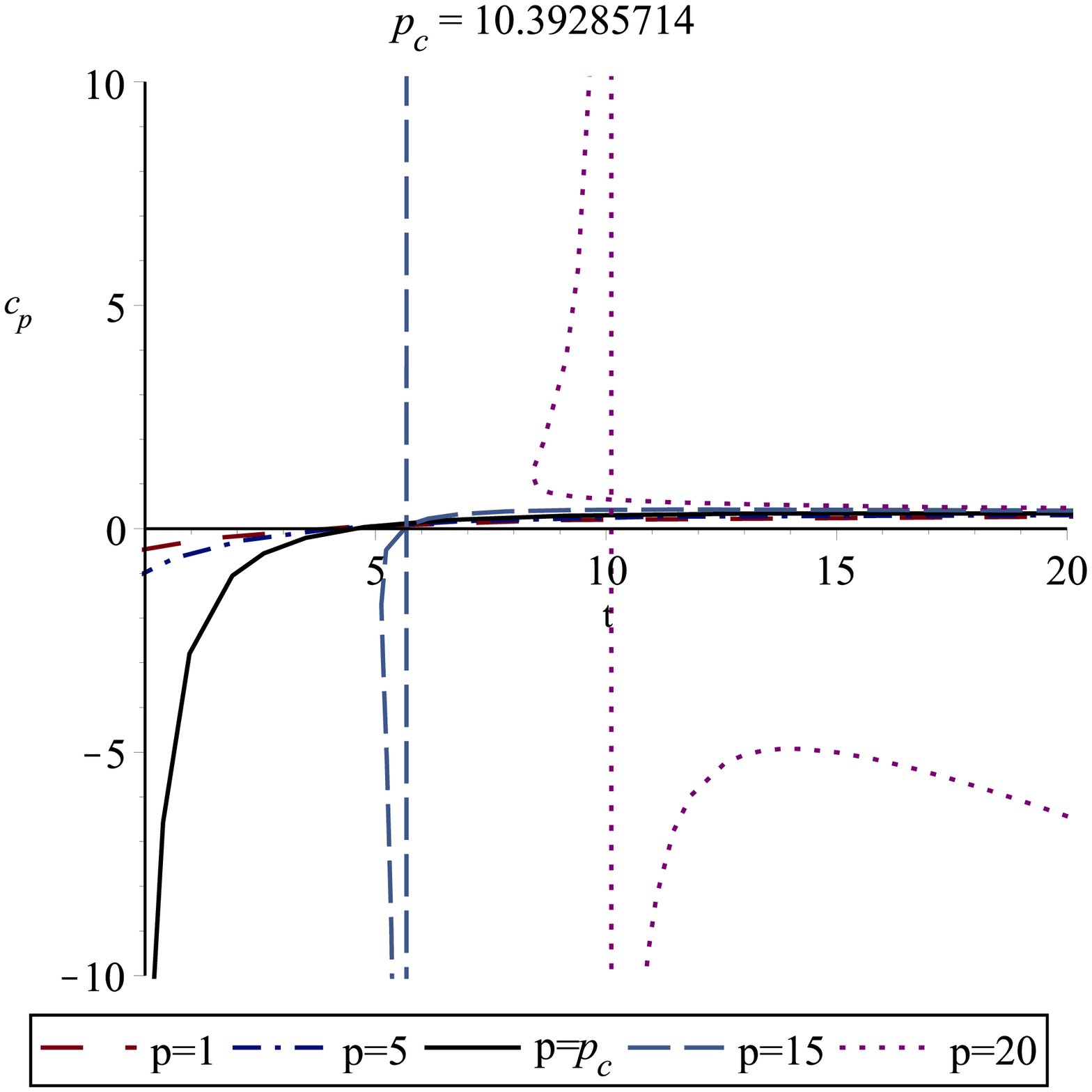}}
 \hspace{2mm}
 \caption{\footnotesize{ (a) Heat capacity curves vs $v$ and (b) Heat capacity vs $t$ at constant pressures given in the Table 1}}
 \end{figure}
\subsection{Phase transition in case of
$\tau\geq1$}Substituting (\ref{state3}) and solving the equations
of critical point (\ref{CE}) we find parametric critical points
for large black hole $\tau\geq1$, such that
\begin{align}\label{critl}&v_c^\pm=\pm i\sqrt{6(1+36\ln\tau)},~~~p_c=\frac{36\ln\tau}{36\ln\tau-1}\bigg[\frac{1}{432\ln\tau(1+36\ln\tau)}-\frac{19}{2\tau^2}\bigg]\notag\\&
t^\pm_c=\frac{\pm
i\sqrt{6(1+36\ln\tau)}}{126\ln\tau(1+36\ln\tau)}.\end{align}
Although the critical points above seem to lack real values in
these regions  apparently, but an examination of  the definition
(\ref{336}) reveals that the  use of $-\infty<\tau<-1$ for large
black holes with negative electric charge $Q<0$ allows us to
obtain real numeric value for the critical points (\ref{critl}).
Numerical studies indicate that for small modified RN-AdS black
hole, i.e. $\tau<1$, the positive values of $0<\tau_c<1$ result in
complex numbers for critical points, but not for $-1\leq\tau_c<0$.
These results are shown for several numeric values in the Table 1.
Moreover, the critical values of parameters for large modified
RN-AdS black hole, i.e. $\tau>1$ are calculated using the equation
\eqref{critl} and collected in the Table 2 for comparison. For the
particular case $\tau=1$ the critical parameters indicated in the
Table 2 are complex numbers and so are excluded from the Table.
\begin{center}
Table 2: The numerical values of critical points for small
modified RN-AdS black hole.  Negative (positive) values for the
critical specific volume $v_c$ and
    the critical temperatures $t_c$ correspond with negative (positive) electric charge.
    \begin{tabular}{|c|c|c|c|c|c|c|c|c|}
        \hline
        $-\tau_c$ &  $-\tilde{\alpha}_c^+$ & $-p_c$ & $\mp v_c^\pm$ & $\pm t_c^\mp$ &  $-\tilde{\alpha}_c^-$ & $-p_c$ & $ \mp v_c^\pm$ &$\pm t_c^\mp$ \\
        \hline
        1 &0.001& 289.05 & $0.271$ &  $115.44$& 0.0003 & 1157 & $p0.190$ & $ 332.56$\\
        0.9 & 0.002& 295.55& $0.269$&  $116.79$ & 0.0006& 836.87& $0.206$ & $259.74$\\
        0.8&0.003& 309.64& $0.267$&  $120.20$& 0.002& 677.91& $0.218$& $ 220.53$\\
        0.7&0.006& 332.66& $0.263$&  $125.88$& 0.003& 596.61 &$ 0.226$ & $198.83$\\
        0.6&0.012& 367.97& $0.257$&  $134.38$& 0.008& 563.04 &$ 0.230$ & $ 188.37$\\
        0.5&0.027& 422.19& $0.250$&  $146.91$& 0.019& 568.76 &$ 0.230$ & $186.99$\\
        0.4&0.069& 510.32& $0.240$&  $165.87$& 0.056& 620.73 &$ 0.227$ & $ 195.28$\\
        0.3&0.232& 670.01& $0.227$&  $196.73$& 0.203& 752.07 &$ 0.219$ & $ 217.59$\\
        0.2&1.248& 1030.1& $0.208$&  $254.90$& 1.163& 1088.5 &$ 0.204$ & $268.61$\\
        0.1 &21.225& 2414.4& $0.178$& $409.36$&20.729&2450.5 & $ 0.176$ & $ 416.69$\\
        \hline
    \end{tabular}
   \end{center} It is easy to show that in the limit of the largest black hole $\tau \to \infty$, the equation
\eqref{alphac} tends to $\tilde{\alpha}_c^-\to -0.017460 $ and
$\tilde{\alpha}_c^+\to 0$ which are correspond to the perturbed
AdS-RN and unperturbed AdS-RN black holes respectively. Thus, we
proceed with our investigation by using diagrams representing the
thermodynamic parameters within the phase space for a particular
value $\tilde{\alpha}_c^-$ and subsequently, compare our results
with an unperturbed AdS-RN black hole phase transition. By
substituting $\tilde{\alpha}_c^-\approx -0.017460 $ into
\eqref{alphac}, the critical value of $\tau_c\approx-0.5$ is
obtained, which
  satisfying the condition of $\tau<1$. The numerical values of thermodynamic parameters are obtained using \eqref{crit} for $\tilde{\alpha}_c^-=-0.017460.$
  They are collected in Table 3.
  \newpage
 \begin{center}
Table 3: The numerical values of critical points for large
modified RN-AdS black hole. Negative (positive) values for the
critical specific volume $v_c$ and
    the critical temperatures $t_c$ correspond to negative (positive) electric charge.
    \end{center}
    \begin{center}
    \begin{tabular}{|c|c|c|c|c|c|c|c|c|c|c|c|c|}
        \hline
        $\tau_c$ &  $-\tilde{\alpha}_c^+$ & $-p_c$ & $\pm v_c^\pm$ & $\mp t_c^\mp$   &    $-\tilde{\alpha}_c^-$ & $-p_c$&  $\pm v_c^\pm$ & $\mp t_c^\mp$\\
        \hline
        1.5& 0.46&12.18& $0.65$ & $8.16$& 0.66&9.73 & $0.72$& $6.25$\\
        2.0& 0.48&4.74& $0.88$& $3.34$&  0.68& 4.00& $0.96$& $2.57$\\
        2.5& 0.41&2.63& $1.06$& $1.92$&  0.60& 2.26&  $1.16$&  $1.45$\\
        3.0& 0.34&1.69& $1.21$& $1.28$&  0.51& 1.46& $1.34$& $0.94$\\
        3.5& 0.28&1.19& $1.35$& $0.93$& 0.44& 1.02&  $1.51$& $0.66$\\
        4.0& 0.24&0.88& $1.48$& $0.71$& 0.38& 0.76& $1.66$&  $0.49$\\
        4.5& 0.20&0.68& $1.59$& $0.57$& 0.34& 0.58&  $1.81$&  $0.38$\\
        5.0& 0.17&0.54& $1.70$& $0.47$&  0.30& 0.46&  $1.95$& $0.31$\\
        5.5& 0.15&0.44& $1.80$& $0.39$& 0.27& 0.38&  $2.09$& $0.25$\\
        6.0& 0.13&0.37& $1.90$& $0.33$& 0.24& 0.31&  $2.22$& $0.21$\\
        6.5& 0.11&0.31& $1.99$& $0.29$& 0.22& 0.26&  $2.35$& $0.18$\\
        7.0& 0.10&0.27& $2.07$& $0.26$&  0.20& 0.22&  $2.47$& $0.15$\\
        7.5& 0.09&0.23& $2.16$& $0.23$& 0.18&  0.19&  $2.60$&  $0.13$\\
        8.0& 0.078&0.20& $2.24$& $0.21$&  0.17& 0.17&  $2.72$& $0.11$\\
        8.5& 0.07& 0.18& $2.31$& $0.19$& 0.16& 0.15&  $2.83$& $0.10$\\
        9.0&0.06& 0.16& $2.38$& $0.17$&  0.15& 0.13&   $2.95$&  $0.09$\\
        9.5&0.06& 0.14& $2.45$& $0.16$&  0.14& 0.12& $3.06$& $0.08$\\
        10.0&0.05&0.13& $2.52$& $0.14$& 0.13& 0.11&  $3.17$& $0.07$\\
        \hline
    \end{tabular}
   \end{center}
\begin{center}
Table 4: The numerical values of critical points for
$\tilde{\alpha}^-_c=- 0.017460$.
    \begin{tabular}{|c|c|c|c|c|c|}
        \hline
    & $\tau_c$ & $p_c$ & $v_c^\pm$ &  $t_c^\mp$ \\
        \hline
    $\tau<1$ &  $\approx-0.5$ & $-625.10$ & $\mp 0.23$ &  $\pm201.64$  \\
    \hline
    $\tau\geqslant1$ & $\approx1.0$ & $-1102.52$ & $\pm0.19$ & $\mp320.68$ \\
\hline
    \end{tabular}
 \end{center}
Now, in order to study the possible thermodynamic phase
transitions of the  black hole under consideration, we rewrite its
equation of state versus the dimensionless thermodynamic
parameters, such that
 \begin{equation}\label{cp}
    \bar{p}=\frac{p}{p_c},~~~\bar{t}=\frac{t}{t_c},~~~\bar{v}=\frac{v}{v_c},~~~\bar{\alpha}=\frac{\tilde{\alpha}}{\tilde{\alpha}_c},~~~\bar{\tau}=\frac{\tau}{\tau_c}.
 \end{equation}
for which \eqref{tem2} reads
 \begin{subequations}
 \begin{align}
    &\bar{p}= \left(\frac{t_c}{p_c v_c}\right) \frac{\bar{t}}{\bar{v}}-\left(\frac{9}{p_c \tau_c^{2}}\right) \frac{1}{\bar{\tau}^{2}}+\left(\frac{1}{p_c v_c^{4}}\right) \frac{1}{\bar{v}^{4}}-\left(\frac{\tilde{\alpha}_c \tau_c^{5}}{p_c v_c^{8}}\right) \frac{\bar{\alpha} \bar{\tau}^{5}}{\bar{v}^{8}},~~~\tau<1\\
    &\bar{p}= \left(\frac{t_c}{p_c v_c}\right) \frac{\bar{t}}{\bar{v}}-\left(\frac{9}{p_c \tau_c^{2}}\right) \frac{1}{\bar{\tau}^{2}}+\left(\frac{1}{p_c v_c^{4}}\right) \frac{1}{\bar{v}^{4}}+\left(\frac{5\tilde{\alpha}_c \tau_c^{4}}{297p_c v_c^{8}}\right) \frac{\bar{\alpha} \bar{\tau}^{4}}{\bar{v}^{8}},~~~\tau\geq 1
 \end{align}
\end{subequations}
 \begin{figure} \centering
    \subfigure[{}]{\label{fig2a}
        \includegraphics[width=0.45\textwidth]{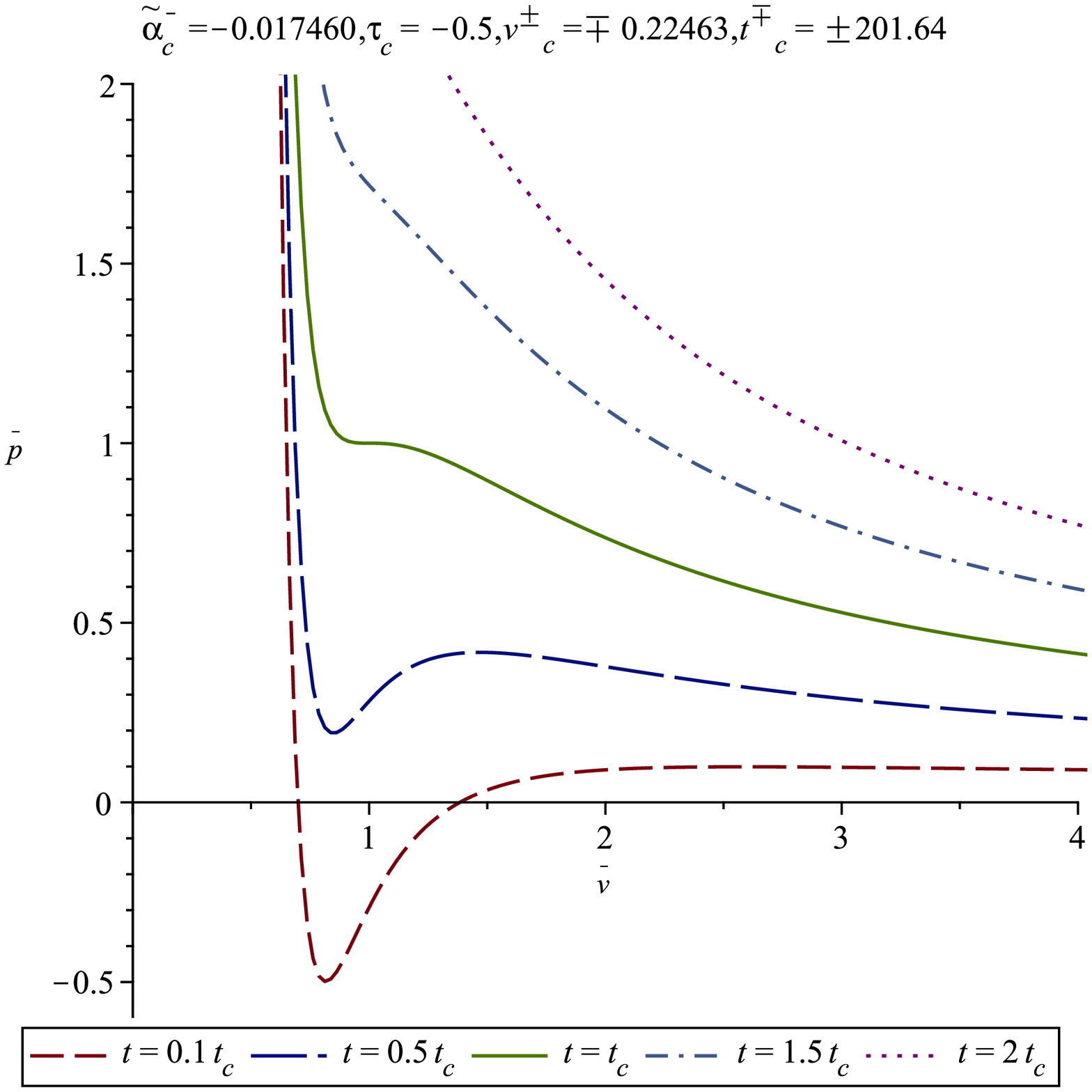}}
    \hspace{2mm}\subfigure[{}]{\label{fig2b}
        \includegraphics[width=0.45\textwidth]{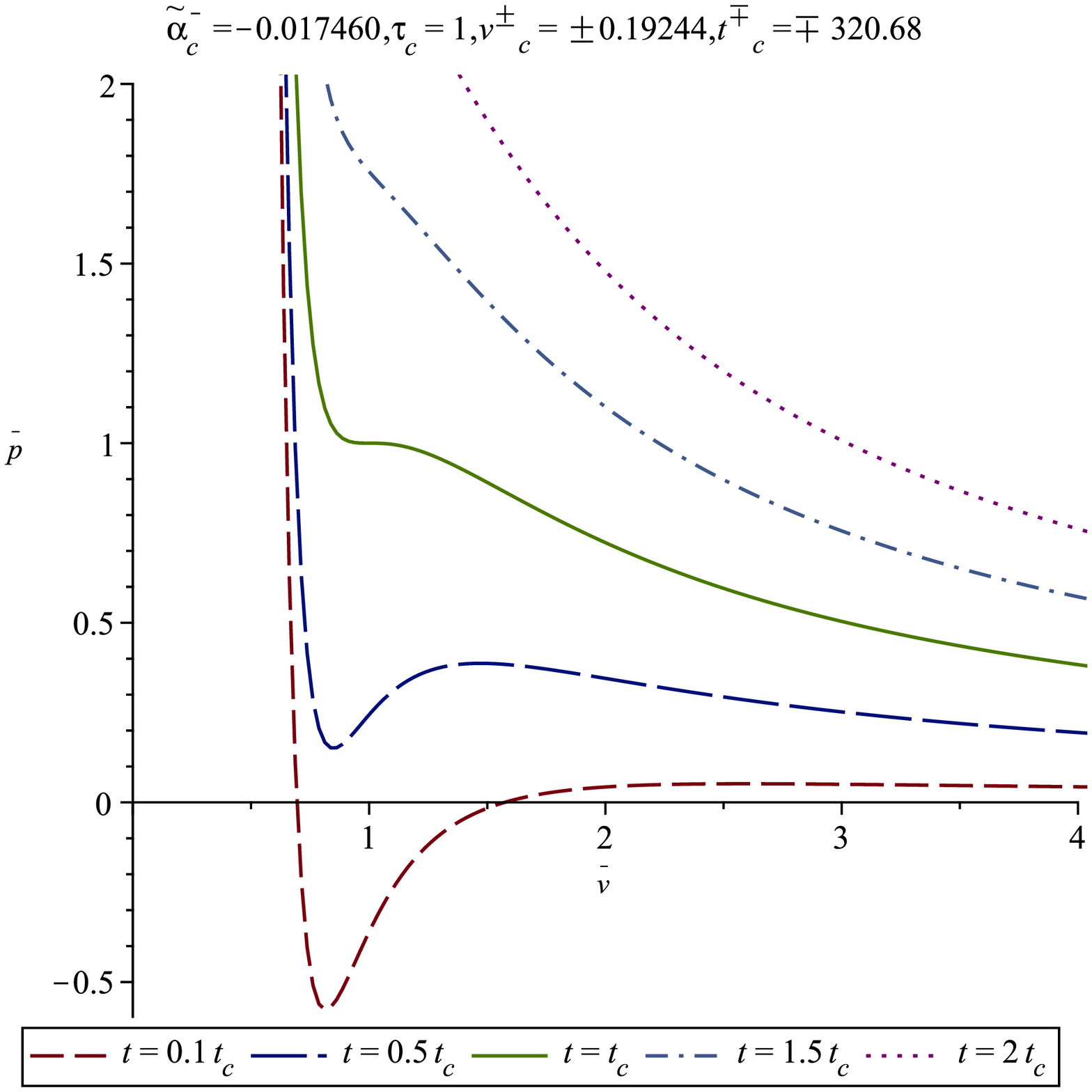}}
    \hspace{2mm}
    \caption{\footnotesize{The isothermal p-v curves for small (a) and large (b) modified RN-AdS black hole according to critical values listed in  the Table 3.}}
 \end{figure}
It can be  seen that for $\tilde{\alpha}_c^-=0.017460$ two values
of $\tau_c\approx-0.5$ and $\tau_c\approx-1$ are obtained, for
which the variation of pressure vs the specific volume  are
depicted for some constant temperatures in figures \ref{fig2a} and
\ref{fig2b}, respectively. As can be observed, the increase in
temperature from bottom to top turns the extremum points into the
turning points. For temperatures less than critical one, the black
hole shows multiple phases which is representative of small/large
black hole phase transition. Moreover, for the least temperature,
a phase transition from dS to AdS space occurs. The Hawking-Page
phase transition is observed for large volume, as well. As a
result, the modified RN-AdS black hole shows the Van der Waal-like
behavior similar to that observed in the  RN-AdS black hole which
is shown in Figure \ref{fig6a}. In this regard, the negativity of
volume and temperature can be attributed to the  electric charge
which is identified as the source of the black hole. This
phenomenon does not influence the thermodynamic behavior of the
black hole.

Now, we choose some small and large values of $\tau_c$ from the
Tables 1 and 2 to plot the isothermal curves illustrating the
variation in pressure as a function of the specific volume. They
are illustrated  in Figures 6 and 7.
 \begin{figure} \centering
    \subfigure[{}]{\label{fig3a}
        \includegraphics[width=0.45\textwidth]{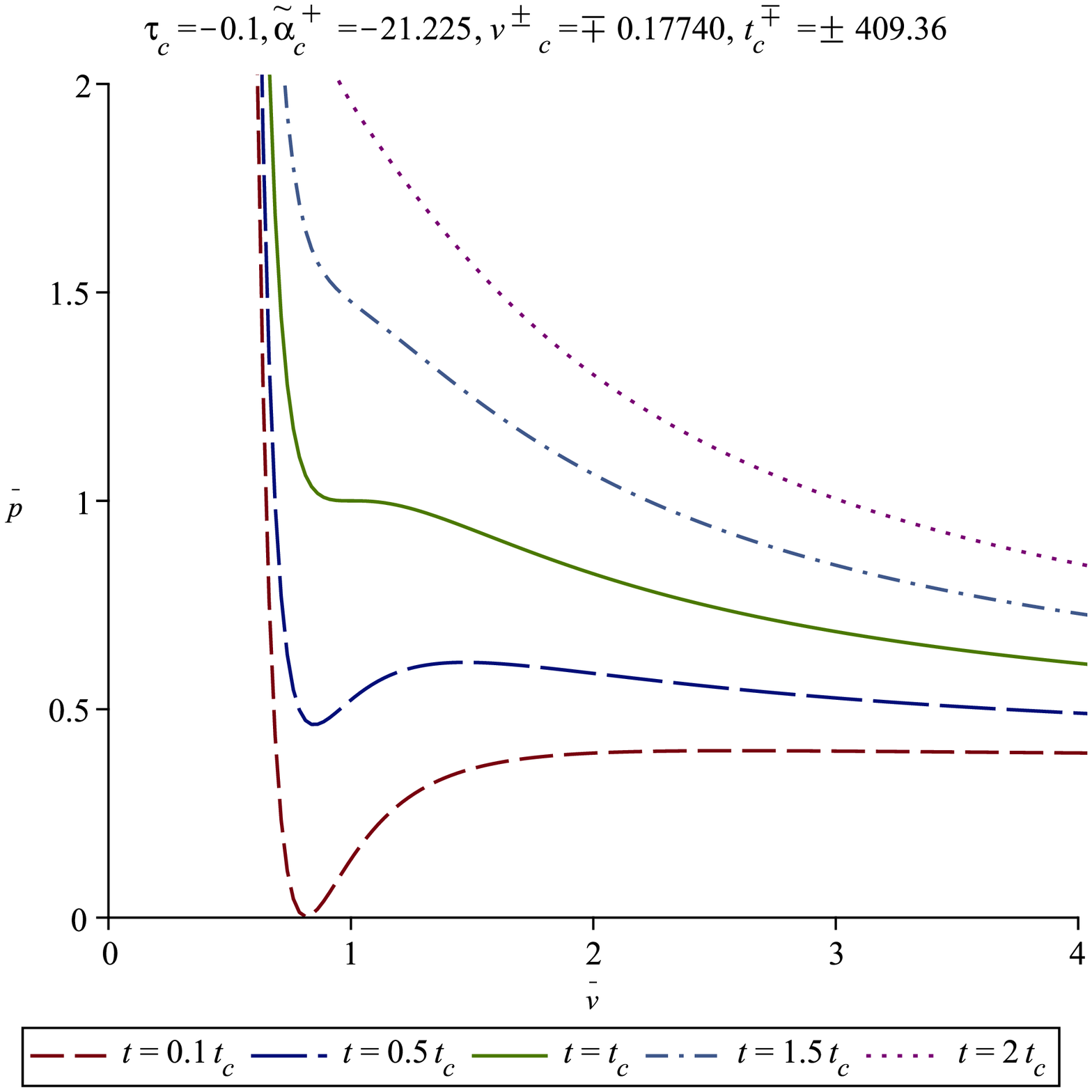}}
    \hspace{2mm}\subfigure[{}]{\label{fig3b}
        \includegraphics[width=0.45\textwidth]{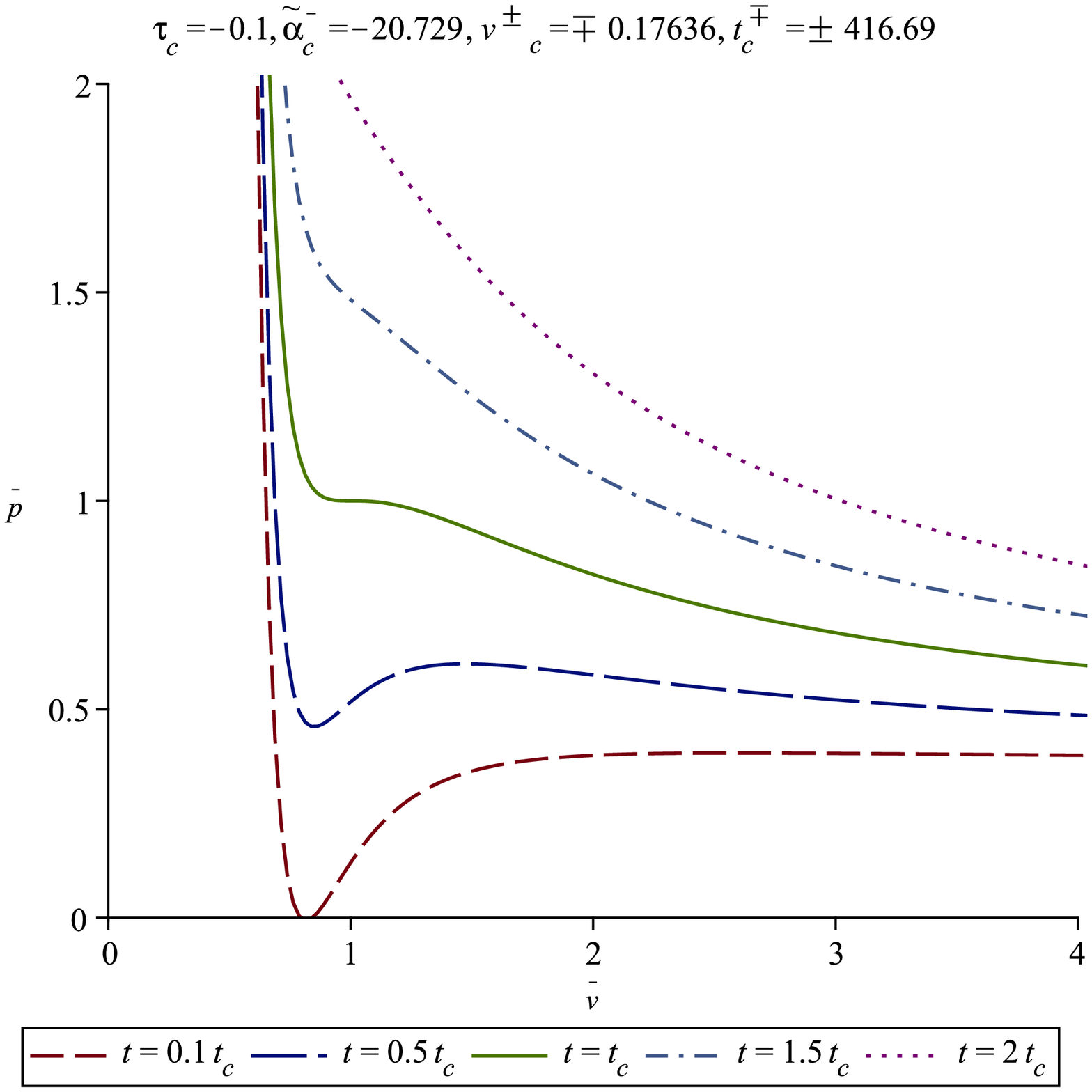}}
    \hspace{2mm}\subfigure[{}]{\label{fig3c}
        \includegraphics[width=0.45\textwidth]{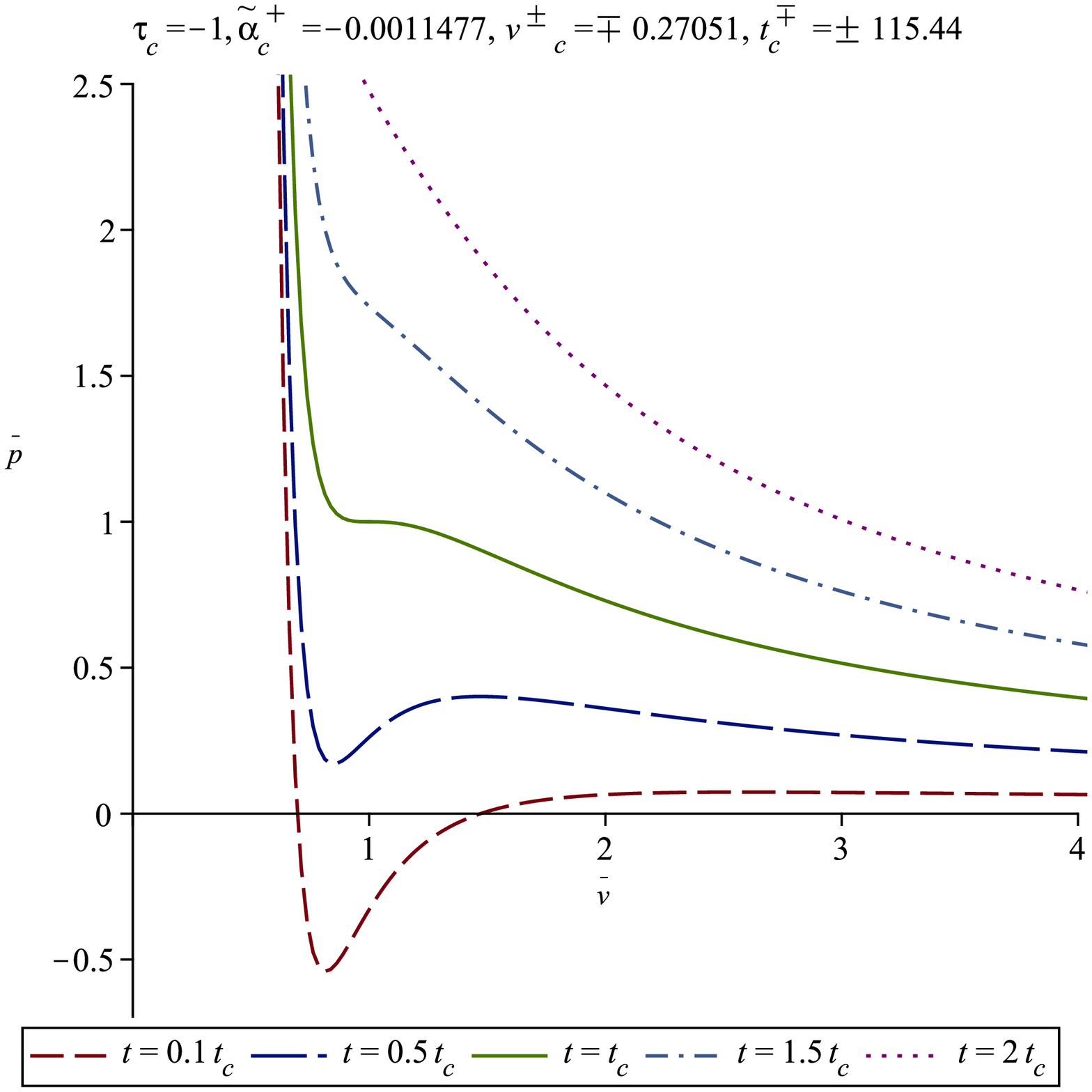}}
    \hspace{2mm}\subfigure[{}]{\label{fig3d}
        \includegraphics[width=0.45\textwidth]{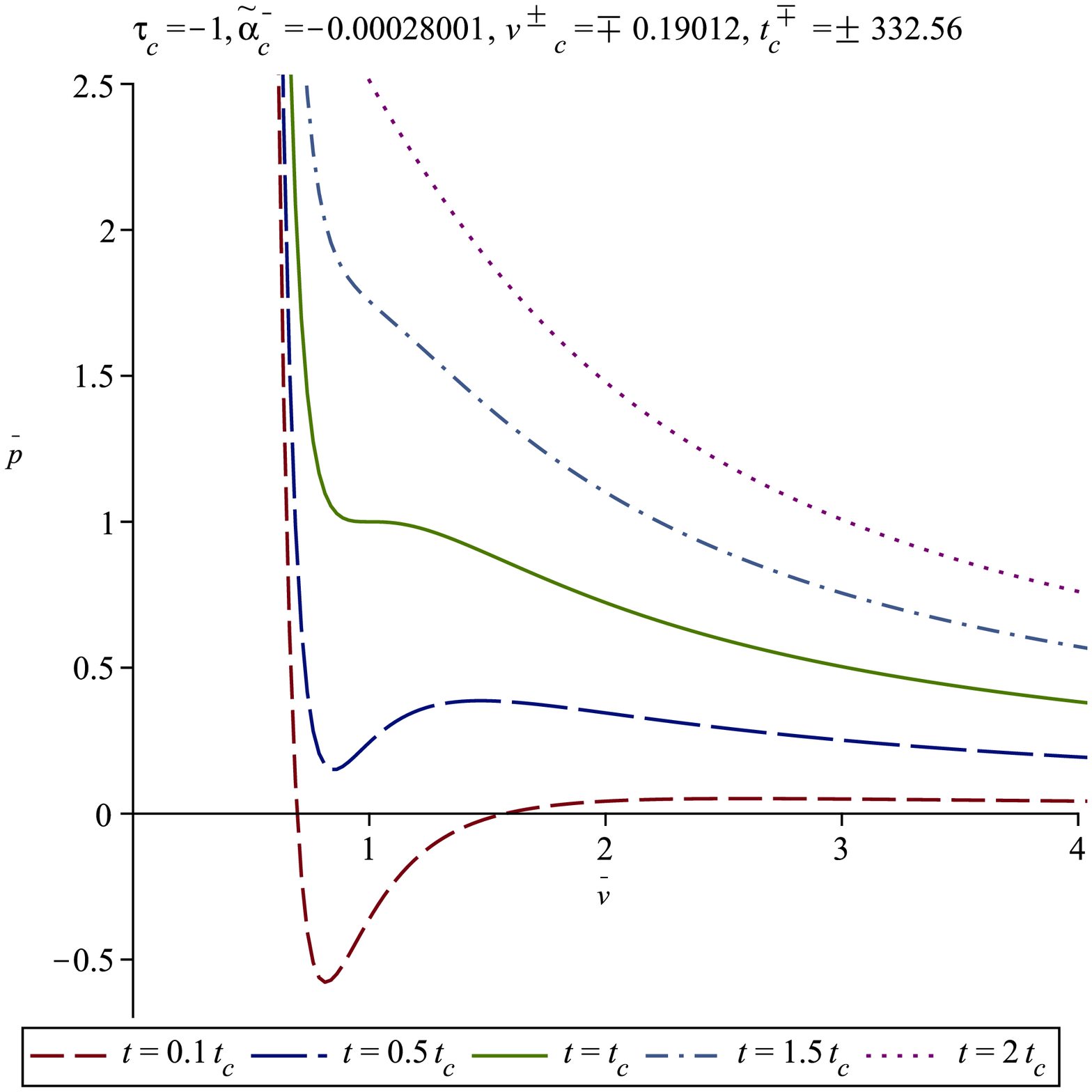}}
    \hspace{2mm}
    \caption{\footnotesize{The isothermal curves of pressure variation in terms of specific volume for small modified RN-AdS black hole for $\tau_c =-0.1$ [\ref{fig3a},\ref{fig3b}] and  $\tau_c =-1$ [\ref{fig3c},\ref{fig3d}].}}
\end{figure}
As one can see, that the small modified RN-AdS black hole shows
the existence of multiple phases and the Van der Waals behavior
for different values of $\tau_c=-0.1$ and $\tau_c=-1$.
Additionally, the Hawking-Page phase transition occurs in large
volumes. For a given value of $\tau_c$, the decrease of absolute
value of the coupling constant $\tilde{\alpha}_c$ reduces the
minimum pressure. As the absolute value of $\tau_c$ increases, in
addition of small/large phase transition, the dS to AdS phase
transition is observed.

 \begin{figure} \centering
    \subfigure[{}]{\label{fig4a}
        \includegraphics[width=0.45\textwidth]{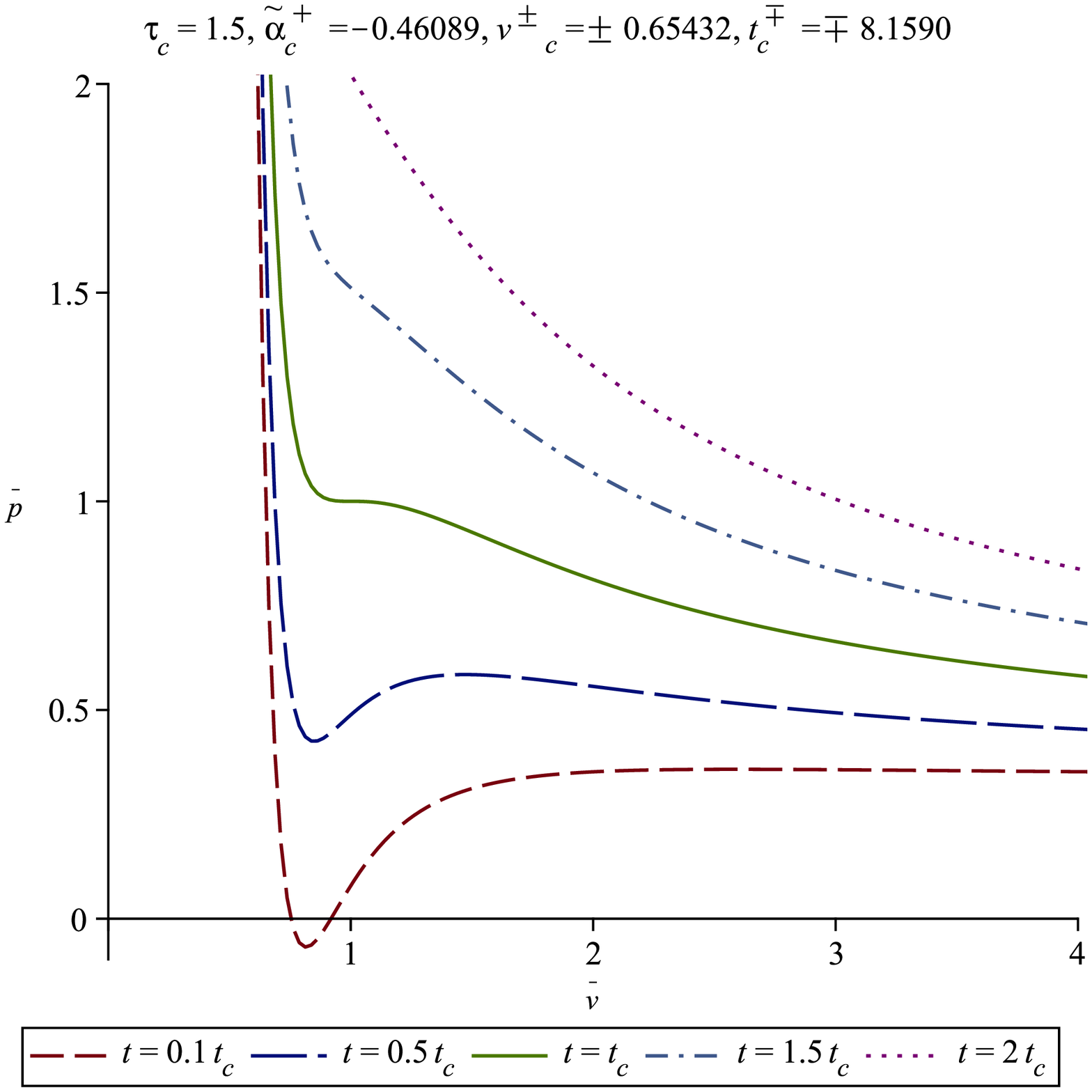}}
    \hspace{2mm}\subfigure[{}]{\label{fig4b}
        \includegraphics[width=0.45\textwidth]{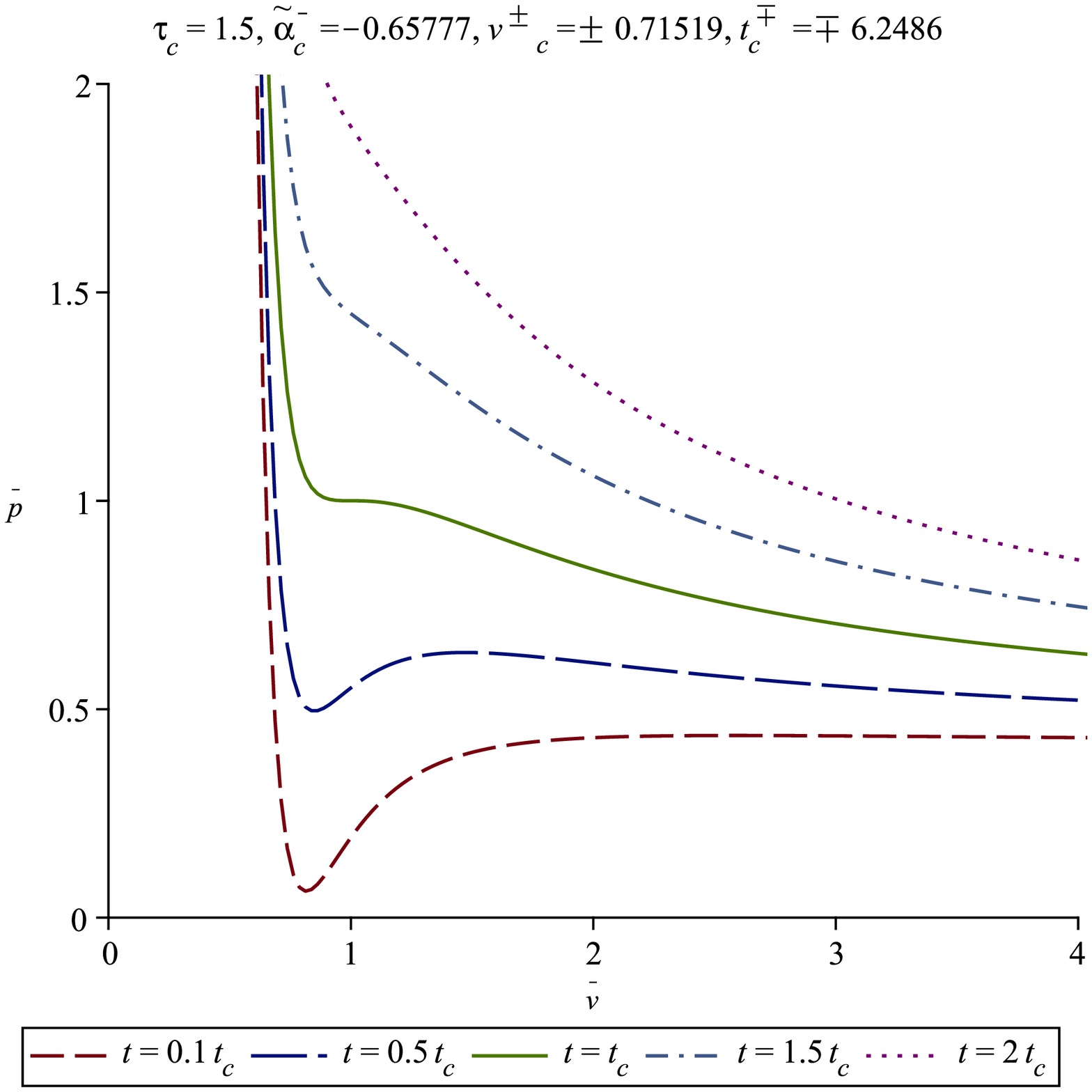}}
    \hspace{2mm}\subfigure[{}]{\label{fig4c}
        \includegraphics[width=0.45\textwidth]{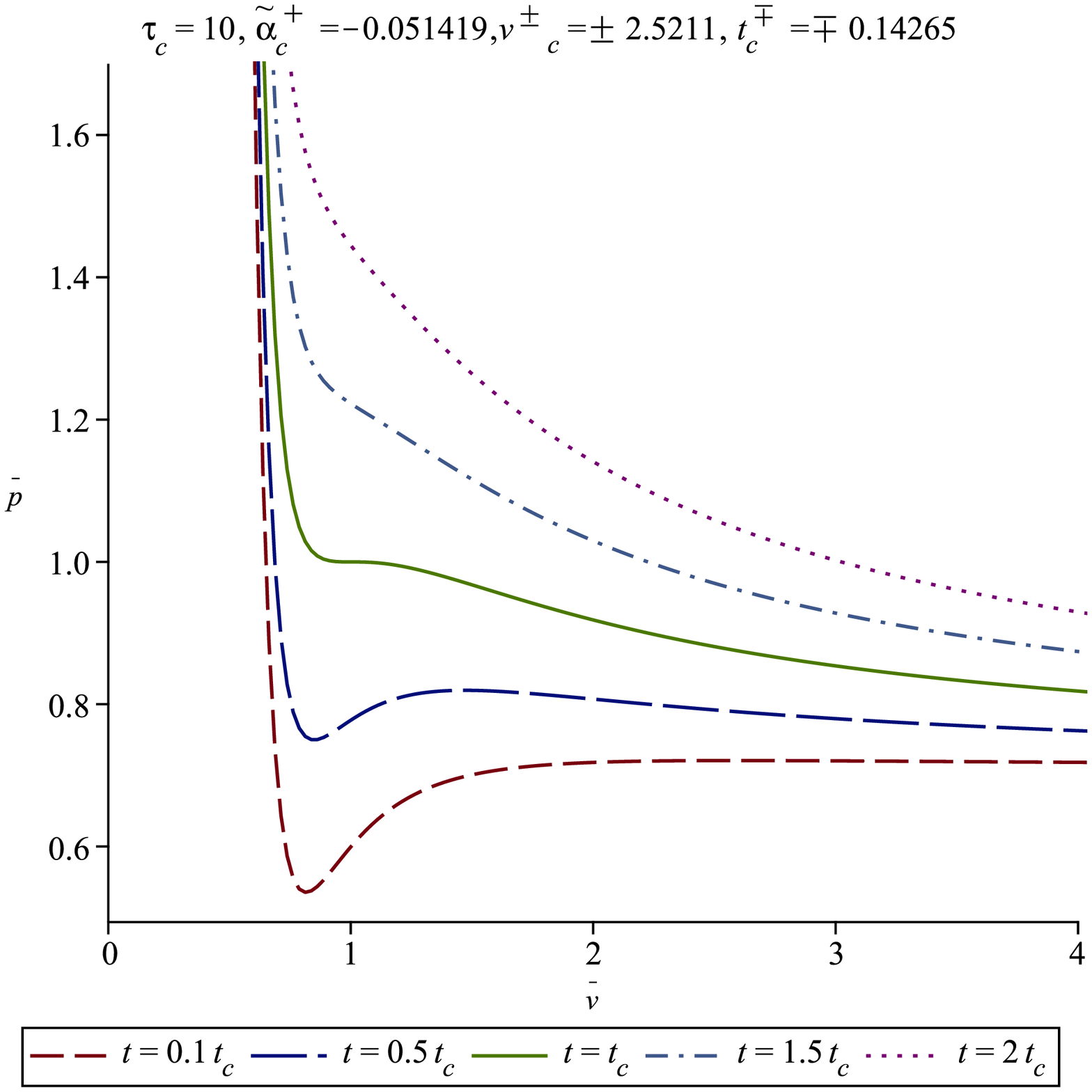}}
    \hspace{2mm}\subfigure[{}]{\label{fig4d}
        \includegraphics[width=0.45\textwidth]{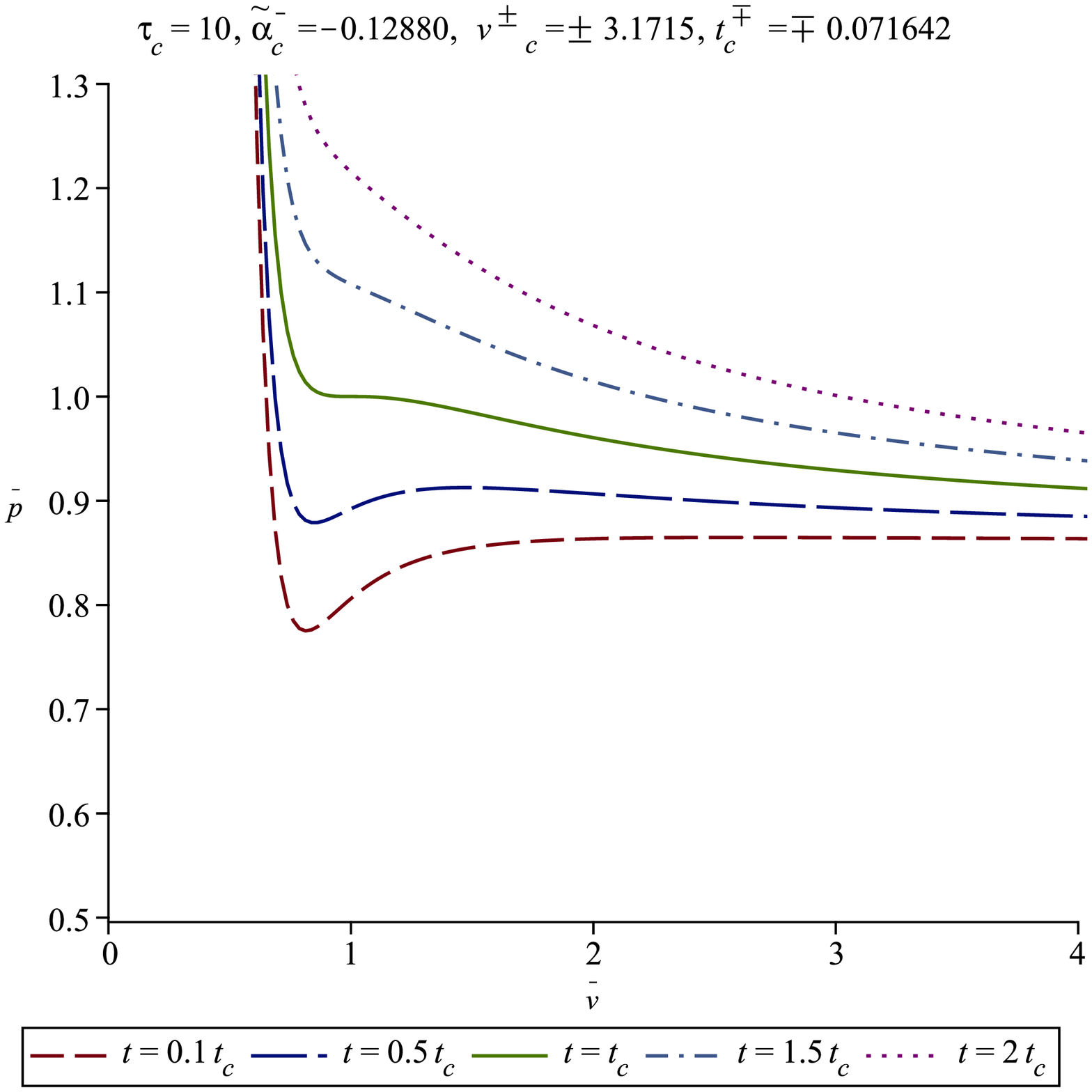}}
    \hspace{2mm}
    \caption{\footnotesize{The isothermal curves of pressure variation in terms of specific volume for large modified RN-AdS black hole for $\tau_c =1.5$ [\ref{fig4a},\ref{fig4b}] and  $\tau_c =10$ [\ref{fig4c},\ref{fig4d}].}}
\end{figure}

In the case of a large modified RN-AdS black hole, the multiple
phases, namely, the Van der Waals behavior and the Hawking-Page
phase transition for different values of $\tau_c=1.5$ and
$\tau_c=10$ are seen in the Figures 6 and 7. For a given value of
$\tau_c$, the increase of the absolute value of the coupling
constant $\tilde{\alpha}_c$ reduces the minimum pressure.
Furthermore, an increase in the absolute value of $\tau_c$,
results in the disappearance of dS/AdS phase transition.

To gain further insight into the thermodynamics of a modified
RN-AdS black hole, we find the Gibbs free energy using the
relation $G=M-TS$. The Gibbs free energy is employed in the
investigation of the stability/instability of black holes. In this
manner, a dimensionless form of the Bekenstein entropy is defined
by
\begin{equation}
    s=\pi v^2,~~~ s=\frac{S_B}{Q^2}.
\end{equation}
The dimensionless form of mass is obtained from \eqref{mass}
\begin{equation}
    m=\frac{9v^3}{\tau^2},~~~ m=\frac{M}{Q}
\end{equation}
Using above relations and temperature \eqref{tem2}, we can obtain
the dimensionless form of Gibbs free energy by $g=m-ts$ which
yields the following results for small and large modified RN-AdS
black holes, respectively.
\begin{align}\label{gibss1}
    &g_<=-\left[ \left(p+\frac{9}{\tau^{2}}\right)\pi -\frac{9}{\tau^{2}}\right] v^{3}+\frac{\pi}{v}-\frac{\pi  \tilde{\alpha}  \,\tau^{5}}{v^{5}},~~~\tau<1\notag\\&
    g_{\geq}=-\left[ \left(p+\frac{9}{\tau^{2}}\right)\pi -\frac{9}{\tau^{2}}\right] v^{3}+\frac{\pi}{v}+\frac{5 \pi \tilde{\alpha}  \,\tau^{4}}{297 v^{5}},~~~\tau\geq1.
\end{align}
The critical values of the Gibbs free energy are obtained by
substituting the critical parameters \eqref{crit} into the above
equations, such that
\begin{align}
    &g_c^{\pm}=\pm \frac{14^{\frac{3}{4}} 3^{\frac{1}{4}} \left(8 \pi+147 \tilde{\alpha}  \,\tau^{3} \right)}{49 \tau  \left(\tilde{\alpha}  \tau \right)^{\frac{1}{4}}},~~~\tau<1\notag\\&
    g_c^{\pm}=\pm \frac{70^{\frac{3}{4}} 11^{\frac{1}{4}} \left(792 \pi-245 \tilde{\alpha}  \,\tau^{2} \right)}{8085 \left(-\tilde{\alpha} \right)^{\frac{1}{4}} \tau},~~~\tau\geq1.
\end{align}
   \begin{center} Table 5: The numerical values of critical points for small modified RN-AdS black hole.
   \end{center}
   \begin{center}
    \begin{tabular}{|c|c|c|c|c|}
            \hline
            $\tau_c$ &  $\tilde{\alpha}_c^+$  &  $g_c^\pm$& $\tilde{\alpha}_c^-$ & $g_c^\pm$ \\
            \hline
            -1 &-0.0011477 & $\mp 26.724$ &  -0.00028001  & $\mp 37.833$\\
            -0.9 & -0.0019140& $\mp 26.864$& -0.00065925&  $\mp 37.833$ \\
            -0.8&-0.0033188& $\mp 27.171$&  -0.0014776&  $\mp 33.081$ \\
            -0.7&-0.0060845&  $\mp 27.655$& -0.0033075& $\mp 32.031$\\
            -0.6&-0.012053&  $\mp 28.348$& -0.0076830 & $\mp 31.554$\\
            -0.5& -0.026633& $\mp 29.325$&  -0.019307&  $\mp 31.615$\\
            -0.4&-0.069131&  $\mp 30.728$& -0.055611&  $\mp 32.285$\\
            -0.3&-0.23205& $\mp 32.868$& -0.20287& $\mp 33.840$\\
            -0.2&-1.2475& $\mp 36.583$& -1.1633& $\mp 37.089$\\
            -0.1 &-21.225&$\mp 45.502$& -20.729&  $\mp 45.652$\\
            \hline
    \end{tabular}
\end{center}
   \begin{center}
   \newpage
Table 6: The numerical values of critical points for large
modified RN-AdS black hole.
\end{center}
\begin{center}
   \begin{tabular}{|c|c|c|c|c|}
            \hline
            $\tau_c$ &  $\tilde{\alpha}_c^+$ & $g_c^\pm$  &$\tilde{\alpha}_c^-$ &  $g_c^\pm$\\
            \hline
            1.0& complex&  complex& complex&  complex\\
            1.5& -0.46089&   $\pm 12.095$& -0.65777& $\pm 11.504$\\
            2.0& -0.47974&  $\pm 9.6883$&  -0.68024&  $\pm 9.4682$\\
            2.5& -0.41076&  $\pm 8.4897$&  -0.59760&   $\pm 8.4403$\\
            3.0& -0.34062& $\pm 7.7043$&  -0.51230&  $\pm 7.7698$\\
            3.5& -0.28246&  $\pm 7.1259$& -0.44038&  $\pm 7.2835$\\
            4.0& -0.23615&  $\pm 6.6737$& -0.38203&   $\pm 6.9090$\\
            4.5& -0.19947&  $\pm 6.3037$& -0.33497&  $\pm 6.6079$\\
            5.0& -0.17018&  $\pm 5.9925$&  -0.29666& $\pm 6.3601$\\
            5.5& -0.14656&  $\pm 5.7257$& -0.26520&  $\pm 6.1512$\\
            6.0& -0.12732& $\pm 5.4925$& -0.23908&  $\pm 5.9726$\\
            6.5& -0.11143& $\pm 5.2859$& -0.21711&  $\pm 5.8183$\\
            7.0& -0.098203& $\pm 5.1013$&  -0.19848&  $\pm 5.6830$\\
            7.5& -0.087082& $\pm 4.9344$& -0.18254&   $\pm 5.5636$\\
            8.0& -0.077657&  $\pm 4.7833$&  -0.16876&  $\pm 5.4584$\\
            8.5& -0.069599&  $\pm 4.6451$& -0.15678&  $\pm 5.3646$\\
            9.0&-0.062661&  $\pm 4.5175$&  -0.14628&   $\pm 5.2801$\\
            9.5& -0.056653&  $\pm 4.3995$&  -0.13702&  $\pm 5.2036$\\
            10.0& -0.051419& $\pm 4.2906$& -0.12880&  $\pm 5.1356$\\
            \hline
    \end{tabular}
    \end{center}
Using \eqref{cp} and $\bar{g}=\frac{g}{g_c}$, the Gibbs free
energy relations obtained earlier can be transformed into a
dimensionless form, such that
\begingroup\makeatletter\def\f@size{9}\check@mathfonts
    \begin{align}\label{gibss2}
        &\bar{g}=-\left(\frac{v_c^3}{g_c \tau_c^2} \right)\left\lbrace  \left[\left(p_c \tau_c^2 \right) \bar{p}+\frac{9}{\bar{\tau}^{2}}\right]\pi -\frac{9}{\bar{\tau}^{2}}\right\rbrace \bar{v}^{3}+\left( \frac{\pi}{g_cv_c}\right) \frac{1}{\bar{v}}-\left( \frac{\pi  \tilde{\alpha}_c  \,\tau_c^{5}}{g_cv_c^{5}}\right) \frac{\bar{\alpha}  \bar{\tau}^{5}}{\bar{v}^{5}},~~~\tau<1\notag\\&
        \bar{g}=-\left(\frac{v_c^3}{g_c \tau_c^2} \right)\left\lbrace  \left[\left(p_c \tau_c^2 \right) \bar{p}+\frac{9}{\bar{\tau}^{2}}\right]\pi -\frac{9}{\bar{\tau}^{2}}\right\rbrace \bar{v}^{3}+\left( \frac{\pi}{g_cv_c}\right) \frac{1}{\bar{v}}+\left( \frac{5\pi  \tilde{\alpha}_c  \,\tau_c^{4}}{297g_cv_c^{5}}\right) \frac{\bar{\alpha}  \bar{\tau}^{4}}{\bar{v}^{5}},~~~\tau\geq1.
    \end{align}
\endgroup
In the following, we plot diagrams of the Gibbs free energy
\eqref{gibss} vs the temperature \eqref{tem2} at constant
pressures (see Figure 8).

 \begin{figure}\centering
    \subfigure[{}]{\label{fig5a}
        \includegraphics[width=0.45\textwidth]{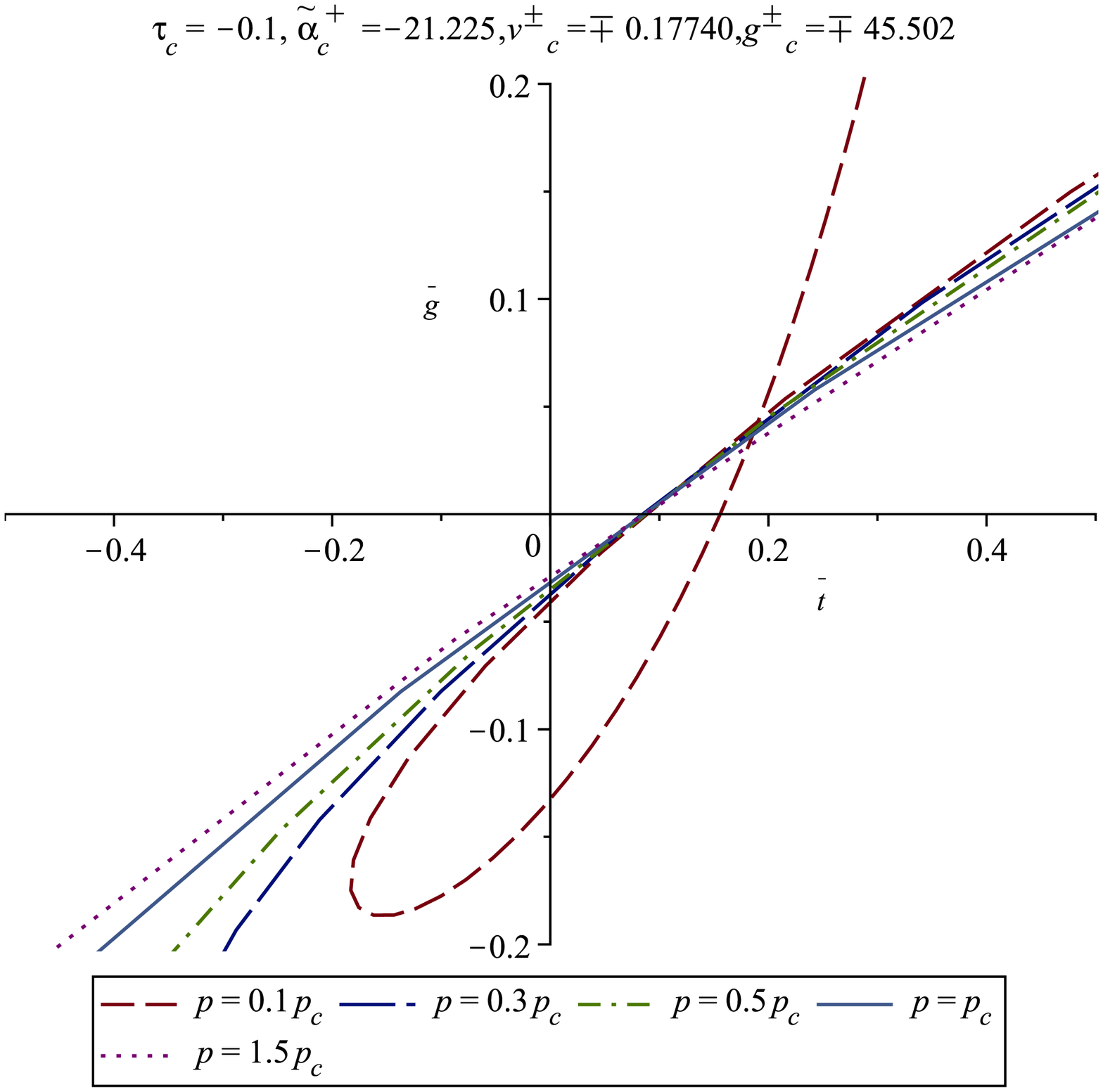}}
    \hspace{2mm}\subfigure[{}]{\label{fig5b}
        \includegraphics[width=0.45\textwidth]{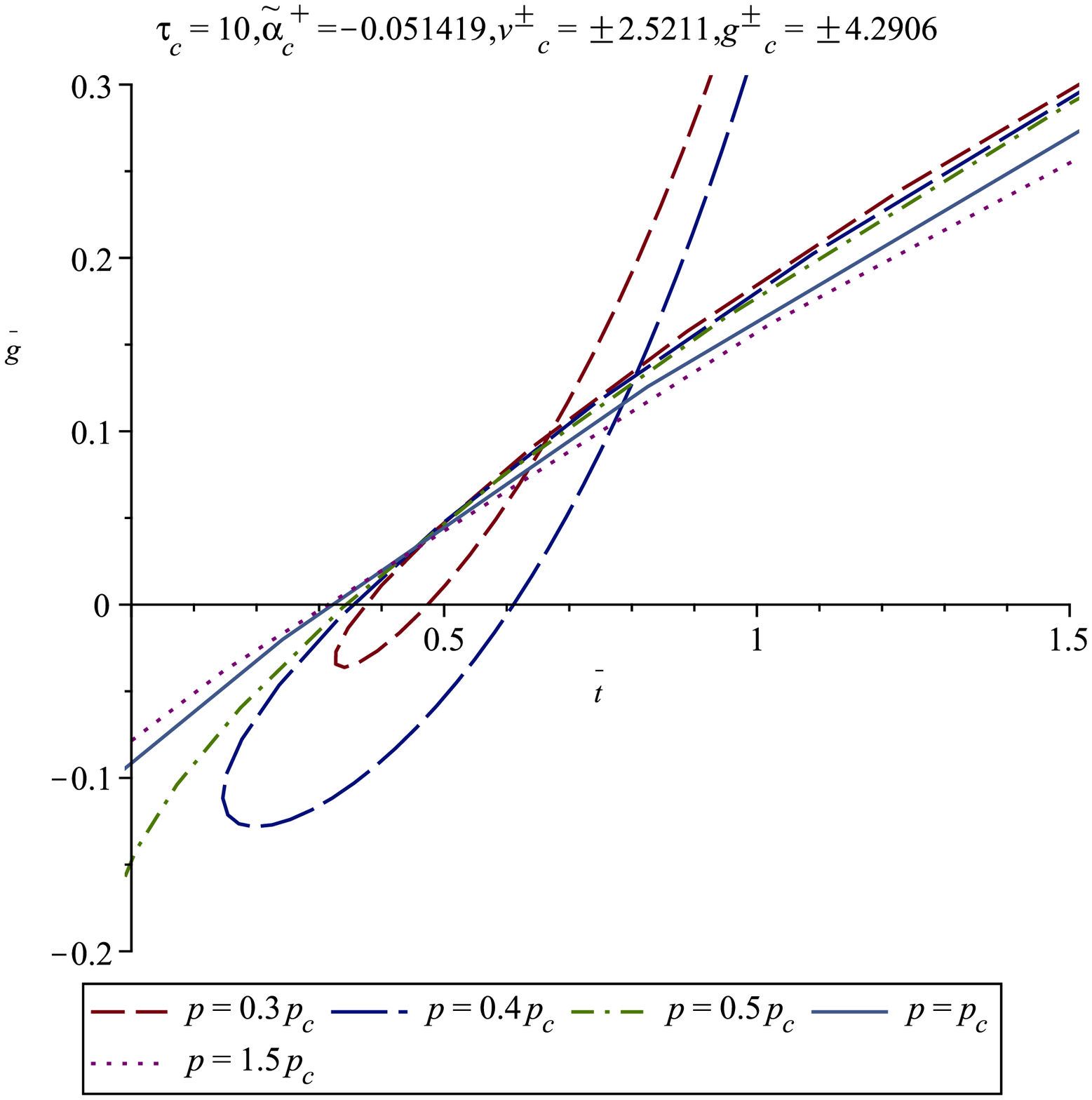}}
    \hspace{2mm}
    \caption{\footnotesize{The isobar curves for g-t corresponding to small [\ref{fig5a}] and large [\ref{fig5b}] modified RN-AdS black holes.}}
\end{figure}

The Diagrams in Figure 8 have been plotted for negative volume,
which is represents a black hole with negative charge according to
the equation \eqref{cp}. Figure \ref{fig5a} is plotted for small
and negative modified RN-AdS black hole. This Figure shows the
swallowtail-like behavior only for the lowest pressure, which is a
coexistence point between multiple phases and a first order phase
transition. The Gibbs free energy behaves similar to the Figure
\ref{fig5a} for other values of $\tilde{\alpha}$ presented in
Table 4 with a slight discrepancy in the minimum temperature of
the curve. The Gibbs free energy variations for negative and large
modified RN-AdS black hole is plotted in Figure \ref{fig5b}. As
one can see, the swallowtail-like behavior occurs at temperatures
less than the critical temperature. The minimum of the curves
exhibits a negative Gibbs free energy which means that the large
modified RN-AdS black hole is in a thermodynamic stable state at
this point.\\A comparison between the results presented in the
previous sections and the thermodynamic behavior of an unperturbed
RN-AdS black hole is a valuable exercise. This is addressed in the
following section.
 \section{RN-AdS black hole}
 The thermodynamic properties of a RN-AdS black hole are
investigated in (n+1)-dimensions by authors of the references
\cite{Chamblin1} and \cite{Chamblin2}. In fact, according to the
findings from preceding sections, our objective is to examine only
the particular case of $\alpha=0.$ In this case, we deal with an
 un-perturbed AdS RN black hole which we should study its
thermodynamic behavior. In the latter case the equation of the
event horizon is obtained by substituting $\alpha=0$ into the
equation \eqref{Hor}, such that
\begin{equation}\label{RNHor}
    1-\frac{2M}{r_+}+\frac{Q^2}{r_+^2}-\frac{\Lambda r_+^2}{3}=0.
\end{equation}
The Hawking temperature of RN-AdS black hole is derived as
follows.
\begin{equation}\label{RNtem}
    T_H=\frac{1}{8\pi}\frac{d   g_{tt}}{dr}\bigg|_{r_+}=\frac{1}{4\pi}\bigg[\frac{M}{r_+^2}-\frac{Q^2}{r_+^3}-\frac{\Lambda r_+}{3}\bigg]
\end{equation}
which can be obtained from \eqref{tem} by setting $\alpha=0$ such
that $T_{<}=T_{>}=T_{RNAdS}$. Because of the black hole
evaporation in the presence of quantum fields interacting with the
horizon yields to be not a constant the black hole mass or
alternatively the enthalpy. In this manner, we must again
eliminate $M$ in the temperature above. This is done by
substituting $M$ from \eqref{RNHor} into \eqref{RNtem}, such that
\begin{equation}
    T_H=\frac{1}{8\pi}\bigg[\frac{1}{r_+}-\frac{Q^2}{r_+^3}-\Lambda r_+\bigg]
\end{equation}
which by using the \eqref{dim} turns to the following
dimensionless form.
\begin{equation}\label{RNtem1}
    t=\frac{3}{2}pv+\frac{1}{2v}-\frac{1}{2v^3}.
\end{equation}
We use the definition of mass
$M=\frac{r_+}{2}+\frac{Q^2}{2r_+}+\frac{8\pi P r_+^3}{3}$ from the
event horizon \eqref{RNHor}  and the Bekenstein entropy $S_B=\pi
r_+^2$ to find the Gibbs free energy through $G=M-TS$, such that
\begin{equation}
    g=\frac{3v}{8}+\frac{5}{8v}+\frac{pv^3}{8},~~~~g=\frac{G}{Q}
\end{equation}
where  we used \eqref{dim}. Now, We able to find the critical
points by solving the critical equations $\frac{\partial
t}{\partial v}|_p=\frac{\partial^2 t}{\partial v^2}|_p=0.$ After
some simple calculations we find
\begin{equation}
    p_c= 0.027778,~~~v_c^\pm=\pm2.4495,~~~ t_c^\pm=\pm0.27216,~~~g_c^\pm=\pm 1.2248.
\end{equation}
Substituting the dimensionless relations
\begin{equation}\label{re}
    \bar{p}=\frac{p}{p_c},~~~\bar{t}=\frac{t}{t_c},~~~\bar{v}=\frac{v}{v_c},~~~\bar{g}=\frac{g}{g_c}
\end{equation} into the temperature relation \eqref{RNtem1}, the dimensionless form
of the equation of state and Gibbs free energy are respectively
\begin{equation}
    \bar{p}=\left(\frac{2t_c}{3p_c v_c} \right) \frac{\bar{t}}{\bar{v}}-\left(\frac{1}{3p_c v_c^2} \right)\frac{1}{\bar{v}^2}+\left(\frac{1}{3p_c v_c^4} \right) \frac{1}{\bar{v}^4},
\end{equation} and
\begin{equation}
    \bar{g}=\left(\frac{3v_c}{8g_c} \right)\bar{v}  +\left(\frac{5}{8v_cg_c}\right) \frac{1}{\bar{v}}+\left(\frac{p_cv_c^3}{8g_c} \right) \bar{p}\bar{v}^3
.\end{equation}

\begin{figure}\centering
    \subfigure[{}]{\label{fig6a}
        \includegraphics[width=0.45\textwidth]{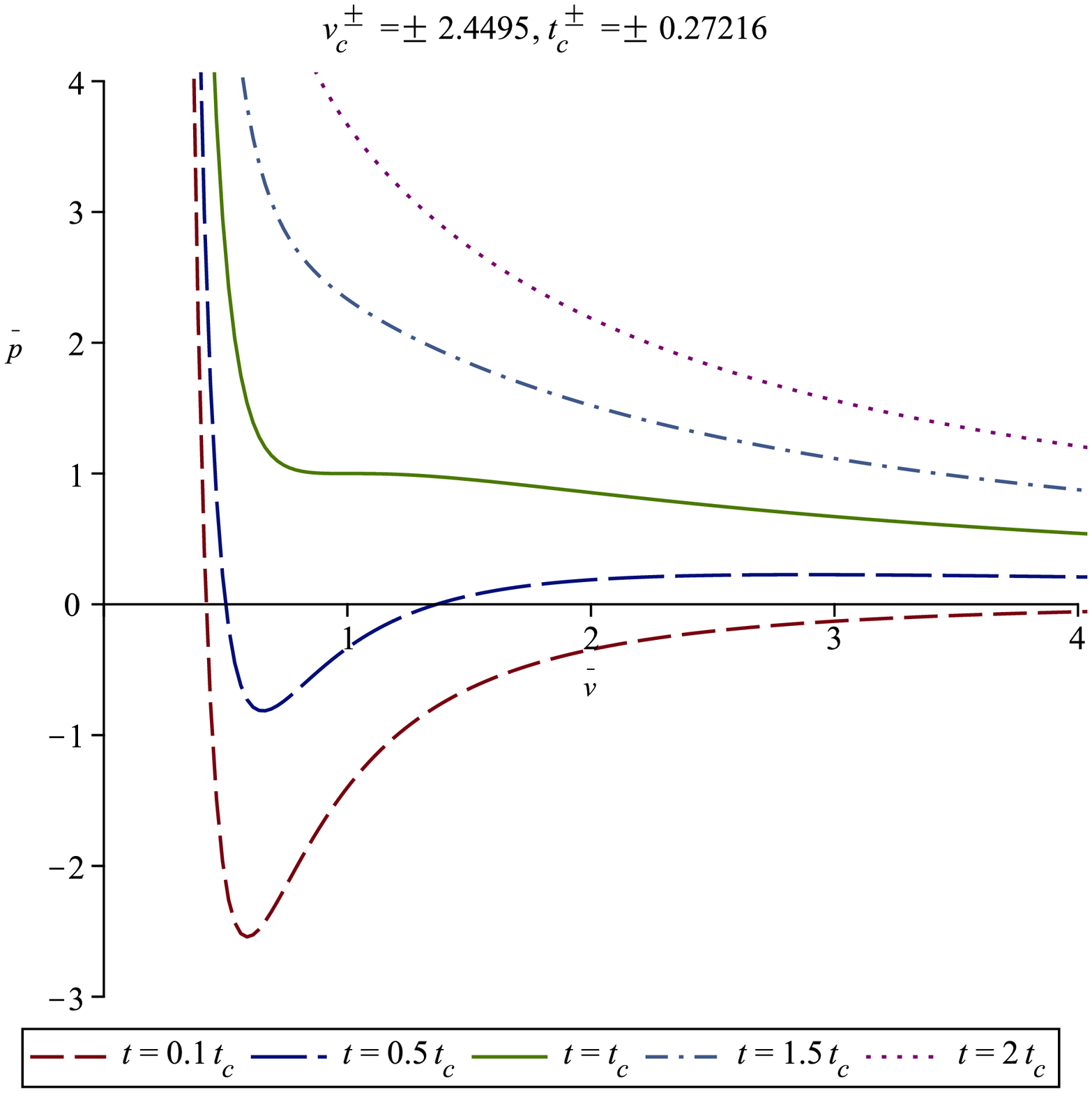}}
    \hspace{2mm}
    \subfigure[{}]{\label{fig6b}
        \includegraphics[width=0.45\textwidth]{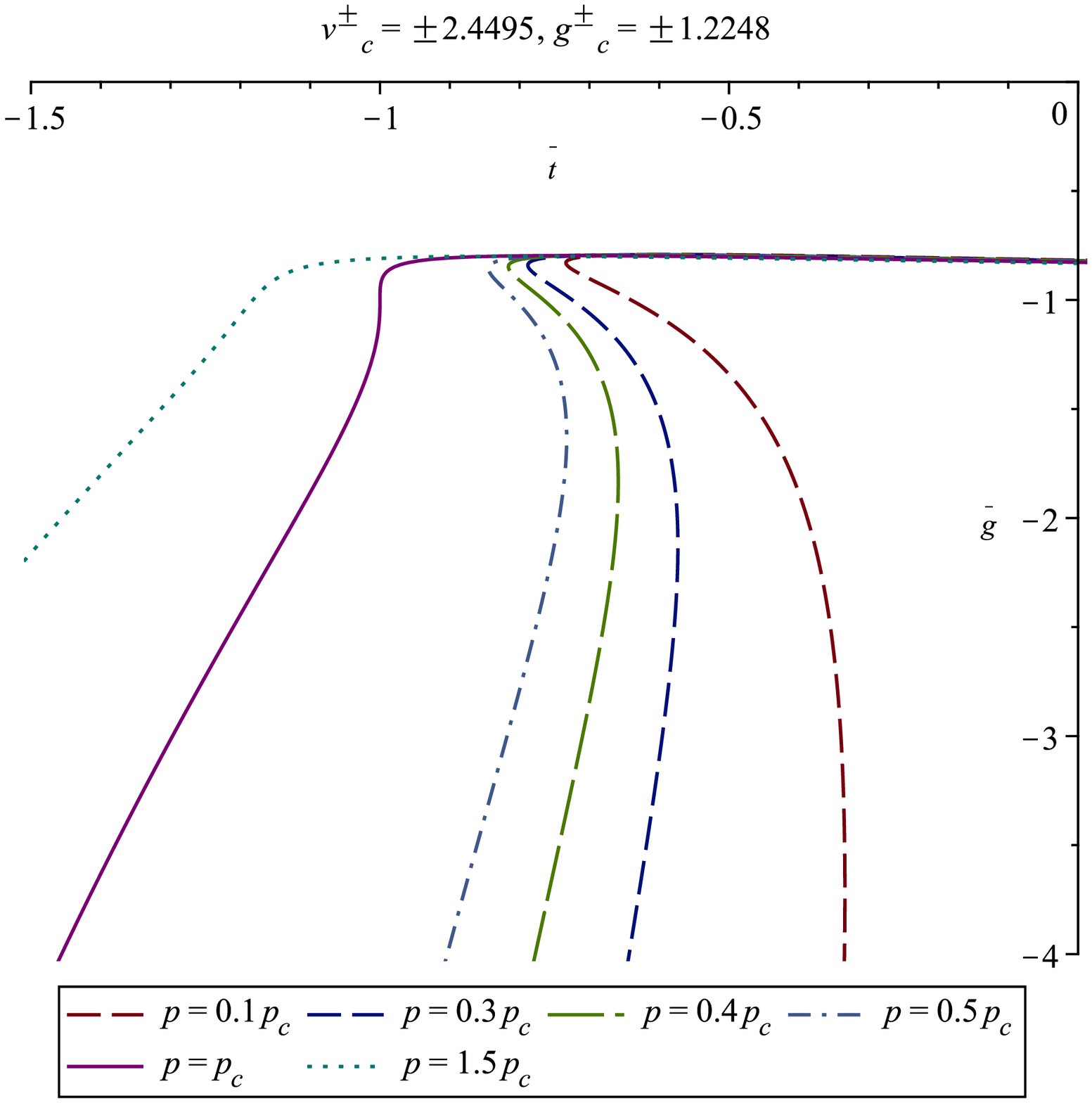}}
    \hspace{2mm}
    \caption{\footnotesize{The isothermal p-v curves and the isobar g-t curves for unperturbed RN-AdS black hole, i.e. $\tilde{\alpha}=0$ and $\tau=1$.}}
\end{figure}

As can be observed in Figure \ref{fig6a}, the unperturbed RN-AdS
black hole shows a Van der Waals behavior for temperatures below
the  critical temperature. In these temperatures, the black hole
shows exitance of a  small/large black hole phase transition,
which is a multi-phase system. For large volume, this black hole
undergoes a Hawking-Page phase transition. By increasing the
pressure from bottom to top, the extremum points become the
turning points and so the RN-AdS black hole behaves same as an
ideal gas. The Figure \ref{fig6b} is plotted for negative volume,
equivalently, for negative charged RN-AdS black hole. This black
hole is in thermodynamic stability for all values of pressures,
while the non-vanishing coupling constant $\alpha$ results in a
first-order phase transition. Comparing the Figure \ref{fig6b}
with [\ref{fig5a}] or [\ref{fig5b}] reveals the significant role
of the interaction parameter $\alpha\neq0$ so that, this
directional interaction between the metric field and the Maxwell
field is responsible for appearance of a coexistence phase between
different phases of the perturbed AdS RN black hole (the loop
curves) which it disappear in the ordinary unperturbed AdS RN
black hole. This constitutes is most significant outcome of this
work.

\section{Concluding remark}
In this study, we employed a gauge invariant modified Einstein
Maxwell gravity to investigate the  thermodynamic behavior of a
extended RN AdS black hole. The motivation for use of this exotic
gravitational theory instead of the well known Einstein Maxwell
gravity is the influence of the cosmic magnetic field which is
present during the cosmic inflationary phase, but it is not
dominated energy in the usual gravity theories. In fact it is
suppressed such that, the vacuum (dark) energy density remains as
the only dominant energy density. The modified part of our used
Lagrangian is a scalar resulting from the contraction between the
symmetric Ricci and the antisymmetric Maxwell tensors. In general.
the metric field equations do not possess exact analytic
solutions, hence we chose the perturbation series method to find
analytic solutions. By determining the horizon equation of this
modified RN AdS black hole we obtained its equation of state via
the calculation of the Hawking temperature. We studied
thermodynamic phase transitions of the black hole. Mathematical
calculations and plotting isobar and isothermal curves in pase
space  for temperature and the pressure as well as  the Gibbs free
energy showed  two different phase transitions so called as the
small/large black hole phase transitions for mini black holes and
the Hawking-Page phase transition for intermediate black holes.
The former phase transitions is similar to the Van der Waals-like
phase transition observed in the ordinary thermodynamic systems.
The positions of the critical points (i.e., the critical pressure,
the critical specific volume and the critical temperature) are
dependent to different values of the interaction parameter between
the electromagnetic and the gravity fields. The results of
mathematical calculations indicated that the phase transitions at
constant pressures below the critical pressure occur at lower
(higher) temperatures with an increase (decrease) in the values of
the interaction parameter. Also at constant temperatures the phase
transitions occur at below the critical temperatures. As the
interaction parameter increases, the extremum pressures, which
determine the type of phase transition, rise, and vice versa.

Furthermore Gibbs free energy diagrams show swallowtail-like
behavior indicating the coexistence of different solid/liquid/gass
phases of the black hole system at a particular point in phase
space. Moreover, a comparison was made between the results
obtained and those of an unperturbed AdS RN black hole. It was
found that the coexistence property disappear in the latter case
however, the same  phase transitions were observed as in the case
of non-vanishing interaction parameter. In this work, we employed
the Bekenstein entropy as the primary and main choice, while the
Hawking proposal regarding the creation of particles/antiparticles
near the horizon causes a modification of the black hole entropy
with a logarithmic term. In the forthcoming work, we intend to
examine the impact of the logarithmic term in the entropy on the
possibility of phase transition for the perturbed AdS RN black
hole under
consideration. Furthermore, Moreover, as a further extension of this work, an investigation for Joule-Thomson expansion of the system under consideration could prove beneficial. \\
\textbf{Funding}\\
 The authors did not receive support from any
organization for the submitted work.\\
 \textbf{Ethics declarations}\\
 \textbf{Conflict of interest}\\
The authors have no competing interests to declare that are
relevant to the content of this article.

\section{Appendixes}
\subsection{Appendix 1}
\begin{align}\label{def}&f_1=\frac{A_0E_0^2}{4}+\frac{2A_0A_0'}{r}+\frac{A_0^2}{r^2}\notag\\&f_2=\frac{3A_0E_0E_0''}{2}+\frac{3A_0E_0'^2}{2}+E_0E_0'A_0'
+\frac{6A_0E_0E'_0}{r}+\frac{E_0^2A'_0}{r}+\frac{3E_0^2A_0}{2r^2}-\frac{E_0^2}{4r^2}\notag\\&
f_3=\frac{E_0^2A_0}{4}+\frac{A_0A'_0}{r}+\frac{A_0^2}{r^2}\notag\\&f_4=\frac{E_0^2}{4A_0}+\frac{A'_0}{rA_0}
\notag\\&f_5=\frac{A_0{E'}_0^2}{2}+\frac{A_0E0E_0''}{2}+\frac{E_0^2A_0''}{2}
-\frac{2E_0^2A_0'^2}{A_0}+\frac{2A_0E_0E'_0}{r}+\frac{E_0^2A'_0}{r}
-\frac{E_0^2A''_0}{2A_0}\notag\\&~~~~
 +\frac{E_0^2A_0}{r^2}-\frac{E_0^2A'_0}{A_0r}-\frac{E_0^2}{4r^2}\notag\\&
 f_6=\frac{A'_0}{4A_0}-\frac{1}{2r}\notag\\&
 f_7=\frac{A_0A'_0}{4}+\frac{A_0^2}{2r}\notag\\&
 f_8=-\frac{E_0^2A_0}{4}+\frac{A_0A''_0}{2} +\frac{{A'}_0^2}{4}+\frac{3A_0A'_0}{2r}\notag\\& f_9=-\frac{E_0^2}{4A_0}+\frac{A''_0}{2A_0}-\frac{{A'}_0^2}{4A_0^2}+\frac{A'_0}{2rA_0}
 \notag\\& f_{10}=\frac{E_0E_1}{2} +{ A_0} {E'}_0^2 +{A_0} {E_0}E''_0
+E_0A'_0E'_0 \notag\\&~~~~+\frac{E_0^2A''_0}{2}
+\frac{4A_0E_0E'_0}{r} +\frac{2E_0^2A'_0}{r}+\frac{2E_0^2A_0}{r^2}
.\end{align}
\subsection{Appendix 2}

\begin{align}&F_1(r)=-\bigg(8\,{\Lambda}^{4}{Q}^{2}+96\,{\Lambda}^{3}
\bigg) {r}^{15}+
 \bigg( 68\,{\Lambda}^{3}{Q}^{2}+576\,{\Lambda}^{2} \bigg)
 {r}^{13}\notag\\&-
 \bigg(108\,{\Lambda}^{3}M{Q}^{2}+1152\,{\Lambda}^{2}M \bigg) {r}^{
12}\notag\\&+
\bigg(42\,{\Lambda}^{3}{Q}^{4}+516\,{\Lambda}^{2}{Q}^{2}-864\,
\Lambda \bigg) {r}^{11}\notag\\&+ \bigg(
612\,{\Lambda}^{2}M{Q}^{2}+3456\, \Lambda\,M \bigg)
{r}^{10}\notag\\&- \bigg(432\,{\Lambda}^{2}{M}^{2}{Q}^{2}+
219\,{\Lambda}^{2}{Q}^{4}+3456\,\Lambda\,{M}^{2}+2088\,\Lambda\,{Q}^{2
} \bigg) {r}^{9}\notag\\&+ \bigg(
279\,{\Lambda}^{2}M{Q}^{4}+3384\,\Lambda\,M{ Q}^{2} \bigg)
{r}^{8}\notag\\&- \bigg(36\,{\Lambda}^{2}{Q}^{6}-1296\,\Lambda
\,{M}^{2}{Q}^{2}+918\,\Lambda\,{Q}^{4}-648\,{Q}^{2} \bigg)
{r}^{7}\notag\\&-
 \bigg(432\,\Lambda\,{M}^{3}{Q}^{2}+630\,\Lambda\,M{Q}^{4}+2160\,M{Q
}^{2} \bigg) {r}^{6}\notag\\&+ \bigg(
108\,\Lambda\,{M}^{2}{Q}^{4}+18\,\Lambda
\,{Q}^{6}+1728\,{M}^{2}{Q}^{2}+1080\,{Q}^{4} \bigg)
{r}^{5}\notag\\&+ \bigg(
162\,\Lambda\,M{Q}^{6}+432\,{M}^{3}{Q}^{2}-2268\,M{Q}^{4} \bigg)
{r}^ {4}\notag\\&-
\bigg(54\,\Lambda\,{Q}^{8}-216\,{M}^{2}{Q}^{4}-702\,{Q}^{6}
 \bigg) {r}^{3}-\bigg( 324\,{M}^{3}{Q}^{4}+378\,M{Q}^{6} \bigg) {r
}^{2}\notag\\&+ \bigg( 324\,{M}^{2}{Q}^{6}+81\,{Q}^{8} \bigg)
r-81\,M{Q}^{8}
\end{align} and \begin{align}&F_2(r)=-88\,{\Lambda}^{4}{r}^{17}-56\,{\Lambda}^{4}{r}^{16}+712\,{\Lambda}^{3
}{r}^{15}- \bigg(1296\,{\Lambda}^{3}M-176\,{\Lambda}^{3} \bigg)
{r} ^{14}\notag\\&+ \bigg(
678\,{\Lambda}^{3}{Q}^{2}-540\,{\Lambda}^{3}M-1992\,{ \Lambda}^{2}
\bigg) {r}^{13}+ \bigg( 558\,{\Lambda}^{3}{Q}^{2}+7056\,
{\Lambda}^{2}M\notag\\&+780\,{\Lambda}^{2} \bigg) {r}^{12}- \bigg(
6048\,{\Lambda}^{2}{M}^{2}+3732\,{\Lambda}^{2}{Q}^{2}+3420\,{\Lambda}^{2}M-
2232\,\Lambda \bigg) {r}^{11}\notag\\&+ \bigg(
6228\,{\Lambda}^{2}M{Q}^{2}+
7344\,{\Lambda}^{2}{M}^{2}+828\,{\Lambda}^{2}{Q}^{2}-11376\,\Lambda\,M
-1872\,\Lambda \bigg) {r}^{10}\notag\\&-
\bigg(1530\,{\Lambda}^{2}{Q}^{4}+
9405\,{\Lambda}^{2}M{Q}^{2}-18144\,\Lambda\,{M}^{2}-6102\,\Lambda\,{Q}
^{2}-8820\,\Lambda\,M\notag\\&+864 \bigg) {r}^{9}+ \bigg(
3744\,{\Lambda}^{2}{
Q}^{4}-8640\,\Lambda\,{M}^{3}-18432\,\Lambda\,M{Q}^{2}-15984\,\Lambda
\,{M}^{2}\notag\\&-2466\,\Lambda\,{Q}^{2}+5616\,M+540 \bigg)
{r}^{8}+\bigg (
11880\,\Lambda\,{M}^{2}{Q}^{2}+4248\,\Lambda\,{Q}^{4}\notag\\&+4752\,\Lambda\,{
M}^{3}+13230\,\Lambda\,M{Q}^{2}-12096\,{M}^{2}-3024\,{Q}^{2}-3132\,M
\bigg) {r}^{7}\notag\\&-
\bigg(4536\,\Lambda\,M{Q}^{4}+108\,\Lambda\,{M}^{2
}{Q}^{2}+3600\,\Lambda\,{Q}^{4}-8640\,{M}^{3}-11340\,M{Q}^{2}\notag\\&-7344\,{M
}^{2}-1674\,{Q}^{2} \bigg) {r}^{6}+ \bigg(
378\,\Lambda\,{Q}^{6}-5994
\,\Lambda\,M{Q}^{4}-7344\,{M}^{2}{Q}^{2}\notag\\&-1998\,{Q}^{4}-4752\,{M}^{3}-
9207\,M{Q}^{2} \bigg) {r}^{5}+ \bigg(
3294\,\Lambda\,{Q}^{6}-6480\,{M
}^{3}{Q}^{2}\notag\\&-1728\,M{Q}^{4}+5832\,{M}^{2}{Q}^{2}+2484\,{Q}^{4}
 \bigg) {r}^{4}+ \bigg( 12312\,{M}^{2}{Q}^{4}+1620\,{Q}^{6}\notag\\&+3564\,{M}
^{3}{Q}^{2}+3186\,M{Q}^{4} \bigg) {r}^{3}-
\bigg(7452\,M{Q}^{6}+11340\,{M}^{2}{Q}^{4}+3564\,{Q}^{6} \bigg)
{r}^{2}\notag\\&+ \bigg( 1458\,{Q}^ {8}+9639\,M{Q}^{6} \bigg)
r-2268\,{Q}^{8}.
\end{align}

\end{document}